\documentclass[sigconf,nonacm,authorversion]{acmart}

\usepackage[nolist]{acronym}
\begin{acronym}

    \acro{3DES}{Triple-DES}
    
    \acro{ACT-R}{Adaptive Control of Thought-Rational}
    \acro{AES}{Advanced Encryption Standard}
    \acro{ALU}{Arithmetic Logic Unit}
    \acro{ANOVA}{Analysis of Variance}
    \acroplural{ANOVA}[ANOVAs]{Analyses of Variance}
    \acro{API}{Application Programming Interface}
    \acro{ARX}{Addition Rotation XOR}
    \acro{ATPG}{Automatic Test Pattern Generation}
    \acro{ASIC}{Application Specific Integrated Circuit}
    \acro{ASIP}{Application Specific Instruction-Set Processor}
    \acro{AS}{Active Serial}
    
    \acro{BDD}{Binary Decision Diagram}
    \acro{BGL}{Boost Graph Library}
    \acro{BNF}{Backus-Naur Form}
    \acro{BRAM}{Block-Ram}
    
    \acro{CBC}{Cipher Block Chaining}
    \acro{CFB}{Cipher Feedback Mode}
    \acro{CFG}{Control Flow Graph}
    \acro{CLB}{Configurable Logic Block}
    \acro{CLI}{Command Line Interface}
    \acro{COFF}{Common Object File Format}
    \acro{CPA}{Correlation Power Analysis}
    \acro{CPU}{Central Processing Unit}
    \acro{CRC}{Cyclic Redundancy Check}
    \acro{CTR}{Counter}
    
    \acro{DC}{Direct Current}
    \acro{DES}{Data Encryption Standard}
    \acro{DFA}{Differential Frequency Analysis}
    \acro{DFT}{Discrete Fourier Transform}
    \acro{DIP}{Distinguishing Input Pattern}
    \acro{DLL}{Dynamic Link Library}
    \acro{DMA}{Direct Memory Access}
    \acro{DNF}{Disjunctive Normal Form}
    \acro{DPA}{Differential Power Analysis}
    \acro{DSO}{Digital Storage Oscilloscope}
    \acro{DSP}{Digital Signal Processing}
    \acro{DUT}{Design Under Test}
    
    \acro{ECB}{Electronic Code Book}
    \acro{ECC}{Elliptic Curve Cryptography}
    \acro{EEPROM}{Electrically Erasable Programmable Read-only Memory}
    \acro{EMA}{Electromagnetic Emanation}
    \acro{EM}{electro-magnetic}
    \acro{EU}{European Union}
    
    \acro{FFT}{Fast Fourier Transformation}
    \acro{FF}{Flip Flop}
    \acro{FI}{Fault Injection}
    \acro{FIR}{Finite Impulse Response}
    \acro{FPGA}{Field Programmable Gate Array}
    \acro{FSM}{Finite State Machine}
    
    \acro{GT}{Grounded Theory}
    \acro{GUI}{Graphical User Interface}
    
    \acro{HCI}{Human Computer Interaction}
    \acro{HDL}{Hardware Description Language}
    \acro{HD}{Hamming Distance}
    \acro{HF}{High Frequency}
	\acro{HRE}{Hardware Reverse Engineering}
    \acro{HSM}{Hardware Security Module}
    \acro{HW}{Hamming Weight}
    
    \acro{IC}{Integrated Circuit}
    \acro{IO}[I/O]{Input/Output}
    \acro{IOB}{Input Output Block}
    \acro{IoT}{Internet of Things}
    \acro{IRB}{Institutional Review Board}
    \acro{IP}{Intellectual Property}
    \acro{IQ}{Intelligence Quotient}
    \acro{ISA}{Instruction Set Architecture}
    \acro{ISCED}{International Standard Classification of Education}
    \acro{IV}{Initialization Vector}
    
    \acro{JTAG}{Joint Test Action Group}
    
    \acro{KAT}{Known Answer Test}
    
    \acro{LFSR}{Linear Feedback Shift Register}
    \acro{LSB}{Least Significant Bit}
    \acro{LUT}{Look-up table}
    
    \acro{MAC}{Message Authentication Code}
    \acro{MAD}{Median Absolute Deviation}
    \acro{MIPS}{Microprocessor without Interlocked Pipeline Stages}
    \acro{MMIO}{Memory Mapped \acl{IO}}
    \acro{MSB}{Most Significant Bit}
    
    \acro{NASA}{National Aeronautics and Space Administration}
    \acro{NCT}{Number Connection Task}
    \acro{NSA}{National Security Agency}
    \acro{NVM}{Non-Volatile Memory}
    
    \acro{OFB}{Output Feedback Mode}
    \acro{OISC}{One Instruction Set Computer}
    \acro{ORAM}{Oblivious Random Access Memory}
    \acro{OS}{Operating System}
    
    \acro{PAR}{Place-and-Route}
    \acro{PCB}{Printed Circuit Board}
    \acro{PC}{Personal Computer}
    \acro{PS}{Processing Speed}
	\acro{PR}{Perceptual Reasoning}
    \acro{PUF}{Physically Unclonable Function}
    
    \acro{RISC}{Reduced Instruction Set Computer}
    \acro{RNG}{Random Number Generator}
    \acro{ROBDD}{reduced ordered binary decision diagram}
    \acro{ROM}{Read-Only Memory}
    \acro{ROP}{Return-oriented Programming}
    \acro{RTL}{Register-Transfer Level}
    
    \acro{SAT}{Boolean satisfiability}
    \acro{SEMOBS}{Self-Modifying Bitstreams}
    \acro{SCA}{Side-Channel Analysis}
    \acro{SHA}{Secure Hash Algorithm}
    \acro{SNR}{Signal-to-Noise Ratio}
    \acro{SPA}{Simple Power Analysis}
    \acro{SPI}{Serial Peripheral Interface Bus}
    \acro{SRAM}{Static Random Access Memory}
    \acro{SRE}{Software Reverse Engineering}
	
	\acro{VLSI}{Very-Large-Scale Integration}
	
	\acro{WAIS-IV}{Wechsler Adult Intelligence Scale}
	\acro{WM}{Working Memory}
    
    \acro{TRNG}{True Random Number Generator}
    
    \acro{UART}{Universal Asynchronous Receiver Transmitter}
    \acro{UHF}{Ultra-High Frequency}
    
    \acro{vFPGA}{virtual FPGA}
	\acro{VC}{Verbal Comprehension}
    
    \acro{WISC}{Writeable Instruction Set Computer}
    
    \acro{XDL}{Xilinx Description Language}
    \acro{XTS}{XEX-based Tweaked-codebook with ciphertext Stealing}

    \acro{ZVT}{Zahlen-Verbindungs-Test}

\end{acronym}

\usepackage{hyperref}
\usepackage{booktabs}
\usepackage{csquotes}
\usepackage{enumerate}
\usepackage{subcaption}
\usepackage{multirow}
\usepackage{rotating}
\usepackage{siunitx}
\usepackage{pdfpages}
\usepackage[inline]{enumitem}
\usepackage{xspace}

\usepackage{tikz}
\usetikzlibrary{backgrounds}
\usetikzlibrary{positioning}
\usetikzlibrary{fit}
\usetikzlibrary{arrows.meta}

\newcommand{\ie}{i.\,e.}
\newcommand{\eg}{e.\,g.}

\newcommand{\etal}{et~al.\@\xspace}
\newcommand{\reversim}{\textsc{ReverSim}\xspace}
\newcommand*\elide{\textup{[\,\dots]}\xspace}

\newcolumntype{H}{>{\setbox0=\hbox\bgroup}c<{\egroup}@{}} % hack to Hide the columns


\copyrightyear{2025}
\setcopyright{cc}
\setcctype{by}

\begin{document}

\title[\reversim]{\reversim: An Open-Source Environment for the Controlled Study of Human Aspects in Hardware Reverse Engineering}

\author{Steffen Becker}
\orcid{0000-0001-7526-5597}
\affiliation{%
  \institution{Ruhr University Bochum}
  \city{Bochum}
  \country{Germany}
}
\additionalaffiliation{%
  \institution{Max Planck Institute for Security and Privacy}
  \city{Bochum}
  \country{Germany}
}
\email{steffen.becker@rub.de}

\author{René Walendy}
\orcid{0000-0002-5378-3833}
\affiliation{%
  \institution{Ruhr University Bochum}
  \city{Bochum}
  \country{Germany}
}
\authornotemark[1]
\email{rene.walendy@rub.de}

\author{Markus Weber}
\orcid{0000-0001-7775-807X}
\affiliation{%
  \institution{Ruhr University Bochum}
  \city{Bochum}
  \country{Germany}
}
\email{Markus.Weber3@rub.de}

\author{Carina Wiesen}
\orcid{0000-0002-4403-1656}
\affiliation{%
  \institution{Ruhr University Bochum}
  \city{Bochum}
  \country{Germany}
}
\authornotemark[1]
\email{carina.wiesen@rub.de}

\author{Christof Paar}
\orcid{0000-0001-8681-2277}
\affiliation{%
  \institution{Max Planck Institute\linebreak  for Security and Privacy}
  \city{Bochum}
  \country{Germany}
}
\email{christof.paar@mpi-sp.org}

\author{Nikol Rummel}
\orcid{0000-0002-3187-5534}
\affiliation{%
  \institution{Ruhr University Bochum}
  \city{Bochum}
  \country{Germany}
}
\email{nikol.rummel@rub.de}

\renewcommand{\shortauthors}{Becker \etal}

\begin{abstract}
\acf{HRE} is a technique for analyzing integrated circuits.
Experts employ \ac{HRE} for security-critical tasks, like detecting Trojans or intellectual property violations,
relying not only on their experience and customized tools but also on their cognitive abilities.
In this work, we introduce \reversim, a software environment that models key \ac{HRE} subprocesses and integrates standardized cognitive tests.
\reversim enables quantitative studies with easier-to-recruit non-experts to uncover cognitive factors relevant to \ac{HRE}.
We empirically evaluated \reversim in three studies. Semi-structured interviews with 14 \ac{HRE} professionals confirmed its comparability to real-world \ac{HRE} processes. Two online user studies with 170 novices and intermediates revealed effective differentiation of participant performance across a spectrum of difficulties, and correlations between participants' cognitive processing speed and task performance. 
\reversim is available as open-source software, providing a robust platform for controlled experiments to assess cognitive processes in \ac{HRE}, potentially opening new avenues for hardware protection.
\end{abstract}

% Keywords and CCS concepts



\maketitle

% Main content includes go here

\newpage
\section{Introduction}
\label{section:introduction}

Understanding the inner workings of \acfp{IC}, also known as \acf{HRE}, is of great importance for various security-critical tasks such as detecting hardware Trojans, checking for intellectual property infringements, or verifying the integrity of cryptographic hardware implementations~\cite{quadir2016survey}. 
All of these tasks have the common goal of increasing trust in \acp{IC}, which underpin a vast array of digital systems, from traditional computers to autonomous cars and medical implants. 
At the same time, the European Chips Act~\cite{eu2023regulation} and the US~CHIPS and Science Act~\cite{repryan20224346} represent massive investments aimed at bolstering trust in \acp{IC}. 
Despite its technical and political importance, \ac{HRE} remains a relatively poorly understood process~\cite{fyrbiak2017hardware}. 

In particular, because \ac{HRE} processes cannot be fully automated, human interaction is critical to their successful execution~\cite{fyrbiak2017hardware,klix2024stealingmaggiessecrets}.
This situates \ac{HRE} within a broader \ac{HCI} framework, similar to fields such as design engineering and medicine, where experts leverage their cognitive abilities and experience to perform computer-supported problem solving through digital interfaces~\cite{Galantucci2006, le2010medical}.
Recent exploratory studies have begun to investigate relevant sensemaking~\cite{becker2020exploratory} and problem-solving strategies~\cite{wiesen2021anatomy}, and cognitive factors~\cite{becker2020exploratory} in \ac{HRE}.
While these studies offer valuable insights --  such as a potential correlation between \ac{HRE} performance and working memory -- their expressiveness is limited by a methodological challenge that we address through the development of a new research method presented in this paper.

\paragraph{Methodological Challenge.}
Researchers aiming to explore human aspects in \ac{HRE} face the methodological challenge that \ac{HRE} experts are largely unavailable to participate in empirical studies involving large realistic problem settings~\cite{becker2020exploratory,wiesen2021anatomy}. 
Previous studies used an approach based on an \ac{HRE} training that enabled students with relevant background knowledge to acquire a sufficient amount of \ac{HRE} skills and tool usage within 14 weeks~\cite{wiesen2018teaching,wiesen2019promoting}. 
Although this approach allowed for an exploratory investigation of human aspects in \ac{HRE}, the sample size of nine participants was quite limited.
Even if more \ac{HRE} experts were available to participate in a research study, the problem remains that they may be accustomed to different \ac{HRE} tools -- making their performance strongly dependent on the tool used in the study.
That, in turn, would make it difficult to measure their performance under the same conditions.

\paragraph{Methodological Approach.}
To address this challenge, we developed and evaluated \reversim, a structured, simplified environment for studying \ac{HRE}.
The software standardizes and reduces the entry requirements for \ac{HRE} studies, thereby broadening the pool of eligible participants.
The design of \reversim was informed by real-world \ac{HRE} problems and tools, with continuous expert feedback and pilot testing guiding its development.
\reversim enables researchers to precisely define sequences of \ac{HRE} tasks and to capture detailed participant interactions, thus facilitating controlled studies investigating the behavior of hardware reverse engineers.
To allow for the study of cognitive factors, \reversim features an extensible integration for psychometric tests.
In this paper, we demonstrate that \reversim is suitable for investigating both \ac{HRE} problem solving and the cognitive factors that influence it, independent of participants' domain-specific prior knowledge or experience with specific tools.
We provide insights into how cognitive processing speed and growing expertise relate to \ac{HRE} performance.

\paragraph{Main Contributions.}
In summary, we make the following contributions:
\begin{itemize}
    \item \textbf{Development of \reversim}. 
    We created \reversim, a study environment designed to model important aspects of real-world \ac{HRE} problems and integrate well-evaluated cognitive tests.
    We provide access to an online demo version and have released \reversim under an open-source license~\footnote{\url{https://github.com/emsec/ReverSim}, see also~\autoref{appendix:materials}} to facilitate future research on human aspects in \ac{HRE}.
    \item \textbf{Experts Validated \reversim}.
    In-depth feedback from 14 \ac{HRE} professionals confirmed that \reversim effectively simulates key aspects of real-world \ac{HRE}, helped us address usability and task-specific issues, and provided detailed suggestions for enhancing its functionality in future iterations.
    \item \textbf{\reversim is Suitable for Controlled Studies with Non-Experts}.
    In a user study involving 109 participants with minimal domain-specific prior knowledge, we found that (a)~university entrance qualification and basic IT experience are sufficient for meaningful engagement with \reversim, and (b)~the environment effectively differentiates participant performance across a wide range of task difficulties.
    These findings affirm \reversim's suitability for controlled studies, providing a robust environment for evaluating \ac{HRE} strategies and performance in non-expert populations.
    \item \textbf{\reversim Enables the Study of Cognitive Factors}.
    We successfully integrated a standardized cognitive test into \reversim, revealing correlations between cognitive processing speed and \ac{HRE} task performance. 
    This capability positions \reversim as an effective method for studying human aspects in \ac{HRE}. 
    It allows for larger samples in a controlled environment, enabling empirical analysis of how cognitive factors influence the \ac{HRE} process.
    \item \textbf{Comparing Novices and Intermediates in \reversim}.
    We compare the \ac{HRE} performance of novices and intermediates in \reversim and find that the relative performance of intermediates closely mirrors that of novices in simpler tasks, but, interestingly, diverges both positively and negatively in more complex tasks. 
    These results suggest that greater expertise does not always lead to greater efficiency.
\end{itemize}

\paragraph{Envisioned Applications of \reversim.}
We consider \reversim as a framework for studying the human aspects of \ac{HRE} with larger samples in a controlled environment, incorporating standardized cognitive tests to empirically analyze how cognitive factors influence the \ac{HRE} process.
We envision that studies using \reversim will aid security researchers in refining hardware protection mechanisms by examining how obfuscation strategies interact with analysts' cognitive processes, including identifying obfuscation strategies that remain effective even as expertise grows.
Understanding these cognitive factors will allow researchers to design targeted countermeasures that effectively hinder reverse engineering efforts.
Additionally, \reversim may have secondary applications in education and \ac{HRE} skill assessments.

\section{Background}
\label{section:background}
In this section, we present relevant background on technical and human aspects in \ac{HRE} that guides the development of our methodological approach.
We end with an overview of our research design and questions.

\subsection{\acl{HRE}}
\label{section:background:hre}
Reverse engineering is the process of extracting knowledge or design information from anything human-made to comprehend its inner structure~\cite{rekoff1985reverse}.
In the hardware security context, it has several applications:
Security engineers are often forced to perform reverse engineering for failure analysis or to identify counterfeited \acp{IC}, security vulnerabilities, or malicious manipulations such as hardware Trojans~\cite{quadir2016survey,wallat2017darkside,puschner2023red}.
\ac{HRE} is also commonly used in research or for competitor analysis, which is legal in many countries~\cite{fyrbiak2017hardware}.
At the same time, \ac{HRE} is associated with illegitimate actions, such as intellectual property infringement, the weakening of security functions, or the injection of hardware Trojans~\cite{quadir2016survey,fyrbiak2017hardware}.

There are two distinct stages in a full-scale \ac{HRE} process~\cite{azriel2021survey}:
In the first stage, a gate-level netlist\footnote{A gate-level netlist is a circuit diagram of Boolean logic gates and memory elements and their interconnections.} is obtained directly from a physical device or by intercepting design information.
Prior research has shown that experienced specialists can reliably extract netlists containing only few mistakes in the extracted gates and connections, which often need to be corrected in the subsequent sense-making stage~\cite{torrance2009state,quijada2018large}.
Our work focuses on this second, sense-making stage of \ac{HRE}, also called netlist analysis or netlist reverse engineering.\footnote{For simplicity, we use \ac{HRE} and the terms \textit{netlist analysis} or \textit{netlist reverse engineering} synonymously in the remainder of this paper.}
Here, an analyst transforms the netlist into higher levels of abstraction that enable detailed analysis and understanding~\cite{subramanyan2014reverse,fyrbiak2018difficulty,azriel2021survey,klix2024stealingmaggiessecrets}. 
This often involves identifying blocks of interest through module recognition~\cite{chisolm1999understanding,hansen1999unveiling,shi2012extracting,subramanyan2014reverse,albartus2020dana}, extraction of control logic and detailed analysis of Boolean subcircuits~\cite{shi2010highly,meade2016netlist,meade2016gate} also known as \textit{word-level reconstruction}~\cite{klix2024stealingmaggiessecrets}, or fully customized approaches~\cite{thomas2015impact}.

While numerous studies have developed (semi-)automated methods for netlist reverse engineering, these approaches often perform optimally in controlled, white-box environments where netlists are synthesized from well-understood hardware descriptions. 
However, the real-world applicability of these automated tools can be limited.
A recent black-box case study by Klix~\etal~\cite{klix2024stealingmaggiessecrets} on netlist reverse engineering underscores this point.
Although the authors achieved a high degree of automation in their \ac{HRE} processes, they found that manual intervention was crucial even for relatively small combinatorial circuits.
For instance, they report: \enquote{While we automated substantial parts of the word-level reconstruction, manual inspection and correction were inevitably necessary to interpret results and address errors.}
This case study demonstrates that, despite advances in algorithmic approaches, the diversity and variability of -- often not completely error-free -- real-world netlists commonly require human ingenuity and customized computing solutions. 
Therefore, the success of \ac{HRE} depends largely on the experience and cognitive skills of the analyst, underscoring the essential role of human intervention in areas where automated tools fall short.

\subsection{Human Aspects in \acs{HRE}}
\label{section:humanfactors}
A few recent works describe the first important findings about the underlying human aspects in \ac{HRE}. 
Lee and Johnson-Laird performed five experiments in a laboratory setting to analyze how participants without prior domain-specific knowledge solved simple reverse engineering problems based on Boolean algebra~\cite{lee2013theory}. 
Subsequently, Lee and Johnson-Laird defined reverse engineering of Boolean systems as a specific but poorly understood kind of human problem solving in which participants had to determine how the mechanics of a specific system worked, which components influenced relevant outputs, and how strongly the components depend on each other. 
However, as the tasks in their experiments were extremely simple, it is questionable to what extent the results are transferable to real-world \ac{HRE}. 

Subsequent research examined problem-solving processes in realistic \ac{HRE} tasks. 
In an exploratory study, Becker~\etal observed the technical processes of hardware reverse engineers analyzing an unknown gate-level netlist~\cite{becker2020exploratory}.
Based on their observations, the authors postulated a three-phase model, in which reverse engineers apply both manual analyses (\eg, visual search and identification of components in the netlist) and semi-automated steps (\eg, verification of crucial netlist components).
Becker~\etal also provided initial insight into cognitive processes in \ac{HRE}, which are largely independent of \ac{HRE} expertise:
The descriptive data from their study suggests that working memory, a factor of intelligence, may play a role in time-efficient problem solving of \ac{HRE} tasks.
Subsequently, Wiesen~\etal conducted a detailed exploration of the problem-solving processes of eight intermediate reverse engineers and one \ac{HRE} expert~\cite{wiesen2021anatomy}.
Their analysis led to a detailed hierarchical problem-solving model that yielded insights into problem-solving strategies and expertise-related differences. 
The authors conclude that superior performance in solving realistic \ac{HRE} tasks may be a function of both expertise and intelligence~\cite{wiesen2021anatomy}.
A drawback of this particular work is that it relies on in-person lab tests for cognitive assessment, making the procedure demanding and time-consuming not only for participants but also for investigators, and thus limiting sample sizes.
Validating these insights with larger samples by examining performance and cognitive factors in a standardized environment suitable for a broad population is a logical next step.

Becker~\etal have proposed that circuit designs aimed at overloading a human attacker's general cognitive abilities may give rise to novel hardware protection techniques, which they coin \enquote{cognitive obfuscation}~\cite{becker2020exploratory}.
While they provide initial insight into the influence of working memory on \ac{HRE} performance, the impact of other factors such as cognitive processing speed -- the rate at which one can scan and sequence information without making errors \cite{weiss2010theoretical} -- is not yet understood.
Uncovering additional cognitive factors relevant to \ac{HRE} in large non-expert samples and examining their impact with small expert samples may thus increase the design space for cognitive obfuscation.
Regarding expertise, Wiesen~\etal suggest studying the problems circuit obfuscation causes in reverse engineers' strategies, some of which might be avoided by more experienced attackers and some of which might persist~\cite{wiesen2021anatomy}.
A first step in this direction could be the identification of \ac{HRE} tasks for which even early expertise acquisition has a clear quantifiable influence on performance.
To enable such controlled experimental studies, we propose a methodological approach in the form of an \ac{HRE} environment called \reversim, which includes a means for administering standardized, unsupervised psychometric tests.

\subsection{Research Design and Questions}
\label{section:rqs}
To empirically evaluate whether \reversim is indeed suitable for the controlled study of human aspects in \ac{HRE}, we conduct two studies, which are guided by the following research questions
\begin{description}[labelindent=\parindent]
    \item[RQ1] What aspects of real-world netlist reverse engineering does \reversim model, according to experts?
    \item[RQ2] Is \reversim suitable for studying aspects of \ac{HRE} with non-experts?
    \begin{enumerate}[label=\alph*)]
        \item What are the minimum requirements to participate in \reversim?
        \item Can we differentiate between participants' \ac{HRE} performance based on the tasks included in \reversim?
    \end{enumerate}
    \item[RQ3] Can \reversim be used to study cognitive factors related to \ac{HRE} performance?
        \begin{enumerate}[label=\alph*)]
            \item Can we integrate standardized cognitive tests into \reversim?
            \item What insights do they provide about the human aspects of \ac{HRE}?
        \end{enumerate}
\end{description}

To answer RQ1, we conduct and evaluate 14 interviews with \ac{HRE} experts who assess the comparability of \reversim with real-world \ac{HRE} processes in \autoref{section:interview}.
To answer RQ2 and RQ3, we analyze data from a user study with 109 participants with low domain-specific prior knowledge in \autoref{section:pilot}.
Extending our investigation beyond novices, we compare the \ac{HRE} performance of novices and intermediates in \reversim in \autoref{section:intermediate}.

\section{Towards the Controlled Study of Human Aspects in \acl{HRE} with \reversim}
\label{section:simulation}
In this section, we outline the development process and underlying design considerations of \reversim.
We followed two principles when designing \reversim:
\begin{enumerate}
    \item \reversim is intended to model important manual subprocesses of real-world netlist reverse engineering (see \autoref{section:background:hre}) as small tasks in a controlled environment to encourage participants -- the problem solvers -- to apply solution strategies comparable to those in the real world.
    \item \reversim should be suitable for a broad range of \ac{HRE} non-experts, without requiring knowledge of any specific tools used in the \ac{IC} industry.
\end{enumerate}
%\pagebreak
We implemented \reversim as a web application, thus providing a high degree of flexibility for study settings: 
Not only can \reversim be used in a laboratory environment, participants can also interact with it in a fully remote setting only requiring a web browser.
\reversim's client is supported by a central server that provides the tasks and records detailed transcripts of participants' interaction with the environment.
This approach enables central administration and data collection for multiple participants simultaneously.
\autoref{fig:simulation:flow:diagram} contains an overview of the typical flow when using \reversim for controlled studies.
To provide a practical view of the environment, we offer a fully functional version of \reversim online (see \autoref{appendix:materials}).

\begin{figure}
    \centering
    %\small
    \begin{tikzpicture}[node distance=1em]
        \node[draw, dashed, align=left, text width=.95\columnwidth] (cognitest) {
            \textbf{Optional Psychometric Test} \\
            Embeds psychometric tests, such as the number-connection test~\cite{oswald2016zvt}, which measures cognitive processing speed, seamlessly into the environment.
        };
        \node[draw, align=left, text width=.95\columnwidth, below=of cognitest] (tutorial) {
            \textbf{Interactive Tutorial} \\
            Introduces and explains basic elements, interaction mechanics, and the overall objective. Participants can try out individual elements in minimal training circuits, particularly supporting those with little prior \ac{HRE} knowledge.
        };
        \node[draw, align=left, text width=.95\columnwidth, below=of tutorial] (qualification) {
            \textbf{Qualification Phase} \\
            Four tasks of lowest possible difficulty (see \autoref{fig:simulation:basicelements:gameui} for an example) verifying participants' basic understanding of the task objective. Participants need to solve each task with a maximum of two attempts. Successfully completing the qualification allows participants to enter the experiment phase; otherwise, they must revisit the tutorial and retry. It is also possible to repeat the tutorial voluntarily. When doing so, users can navigate freely between the individual elements of the tutorial.
        };
        \node[draw, align=left, text width=.95\columnwidth, below=of qualification] (experiment) {
            \textbf{Experiment Phase} \\
            Set of tasks tailored to the study at hand (see \autoref{subsection:leveldesign}). Tasks can be assigned per participant. This phase supports randomization and optionally a means for participants to skip tasks in case they become stuck.
        };
        \draw[->]
            (cognitest) edge (tutorial)
            (tutorial) edge (qualification)
            (qualification) edge (experiment);
    \end{tikzpicture}
    \caption{Explanation of the typical phases when using \linebreak \reversim for controlled studies.}
    \Description[Flowchart of the phases of ReverSim]{The flowchart visualizes the participant's progression through the four phases of the study setting, which will be visited in the following order: "Optional Psychometric Test", "Interactive Tutorial", "Qualification Phase", and "Experiment Phase". In the first phase, tests like the number-connection test, that measures the cognitive processing speed, can be embedded into the environment, if required to answer the research question. The core concepts of the environment are introduced in the second phase with minimal training circuits. In the third phase, four tasks of the lowest possible difficulty are presented. The participant will be directed back to "Interactive Tutorial" (not visualized in the flow of the diagram), if he/she fails to solve the tasks correctly. If successful, the participant will proceed to the last phase, where a set of tasks is shown, which is tailored to the study.}
    \label{fig:simulation:flow:diagram}
\end{figure}

\paragraph{Development Process.}
\label{subsection:development}

\begin{figure*}[t]
    \centering
    \includegraphics[width=\textwidth]{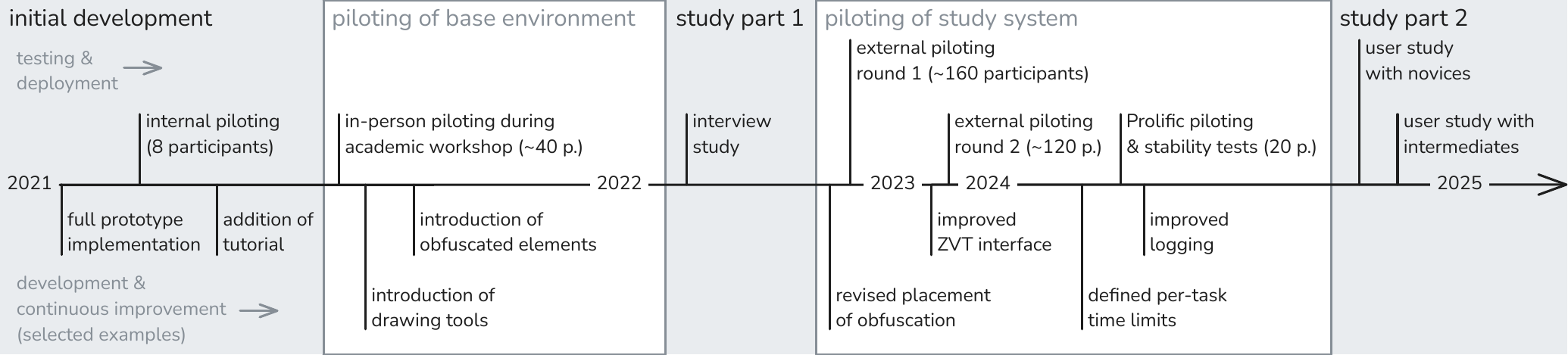}
    \caption{Sketch of the development process of \reversim in terms of testing and deployment (top) as well as major development steps and examples of continuous improvement (bottom). We implemented a full, working prototype of \reversim, followed by a round of internal piloting and feedback from hardware security researchers, \eg, resulting in the addition of the interactive tutorial. Piloting of the resulting base environment took place during an in-person academic workshop with a separate group of \ac{HCI} and security researchers and professionals, \eg, prompting the introduction of the drawing tools. After conducting the interview study (see \autoref{section:interview}), we further revised \reversim and piloted our study setup in two rounds, each with independently recruited participants from industry and academia. We performed a final piloting round with participants from Prolific. Based on the finalized study setup, we conducted two user studies with novices (see \autoref{section:pilot}) and intermediates (see \autoref{appendix:intermediatesample}).}
    \Description[Timeline for the development process of ReverSim]{The timeline visualizes selected improvements that were made to ReverSim over the following five phases: "initial development", "piloting of base environment", "study part 1" (interview study), "piloting of study systems" and "study part 2" (interview studies with novices and intermediates). Most milestones are mentioned in the caption, with the following additions: During "piloting of the base environment," obfuscated elements were introduced after the drawing tools. During the external piloting rounds, the positioning of the obfuscation was revised, the ZVT interface was enhanced, per-task time limits were added, and the logging improved.}
    \label{fig:simulation:development:timeline}
\end{figure*}

\autoref{fig:simulation:development:timeline} illustrates the development process of \reversim between 2021 and 2024, which involved multiple rounds of internal and external piloting.
Feedback from \acs{HCI} and security researchers, as well as computer security conference attendees from industry and academia, informed iterative improvements to the initial prototype.
Examples of changes include the addition of drawing tools and the option to voluntarily retake the interactive tutorial.
As we continue open-source development, we plan to introduce further features that were proposed in this iterative process.
In the following sections, we describe the present version of \reversim.

\subsection{Basic Elements}
\label{subsection:basicelements}
The core of \reversim consists of eight basic elements for Boolean circuits that together form a task:
At all inputs to the circuit, a \textit{battery} provides current and is always connected to a \textit{switch}, which is the only basic element the participant can operate.
The experimenter can define whether each switch is initially open or closed.
Alternatively, \reversim can choose a random starting position.
Through a mouse click, the participant can close and open switches in order to pass or stop the current flow.
Each output is realized by a \textit{lamp} or \textit{danger sign}.

\begin{figure}[t]
    \footnotesize
    \centering
    \begin{tikzpicture}[scale=.7]
        \node[inner sep=0pt,rotate=-90] (bulb_off) at (0,0)
            {\includegraphics[width=0.08\textwidth]{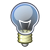}};
        \node[inner sep=0pt,rotate=-90] (bulb_on) at (4,0)
            {\includegraphics[width=0.08\textwidth]{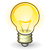}};
        \node[inner sep=0pt,rotate=-90] (danger_off) at (0,2)
            {\includegraphics[width=0.08\textwidth]{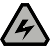}};
        \node[inner sep=0pt,rotate=-90] (danger_on_background) at (4,2)
            {\includegraphics[width=0.12\textwidth]{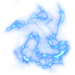}};
        \node[inner sep=0pt,rotate=-90] (danger_on) at (4,2)
            {\includegraphics[width=0.08\textwidth]{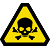}};
        \node[inner sep=0pt] (no_current) at (-1,3.5)
            {no current};
        \node[inner sep=0pt] (no_current) at (3,3.5)
            {current};
        \begin{scope}[on background layer]
            \draw[very thick] (-2, 0) -- (bulb_off.center);
            \draw[very thick,color=yellow] (2, 0) -- (bulb_on.center);
            \draw[very thick] (-2, 2) -- (danger_off.center);
            \draw[very thick,color=yellow] (2, 2) -- (danger_on.center);
        \end{scope}
    \end{tikzpicture}
    \caption{Lamp and danger sign indicate the expected output of a circuit. Participants solve the task by supplying current to the lamp, turning it on (bottom right), and by ensuring that no current is supplied to the danger sign (top left). }
    \Description[lamp and danger sign symbols used in the user interface]{Four symbols are arranged in a two-by-two grid. The top row shows the two variants of the danger sign: On the left, the danger sign is not powered. It is visually represented as a grey triangle with a lightning bolt symbol. On the right, the danger sign is powered. It is visualized as a yellow triangle with a skull and crossbones symbol, surrounded by blue arcing. The bottom row shows the light bulb variants. On the left side, the light bulb not receiving power is not lit. On the right side, the light bulb receiving power is lit.}
    \label{fig:simulation:mechanics:outputs}
\end{figure}

The objective of each task is to determine the correct switch settings to yield the desired circuit output:
For a valid solution, all \textit{lamps} must light up and do so when they are supplied with current.
On the other hand, the \textit{danger signs} must not be supplied with current; otherwise, an electric discharge is displayed at the respective output, indicating an incorrect solution.
\autoref{fig:simulation:mechanics:outputs} illustrates the lamp and danger sign with and without supplied current.

\begin{figure}[t]
    \centering
    \includegraphics[width=.9\columnwidth]{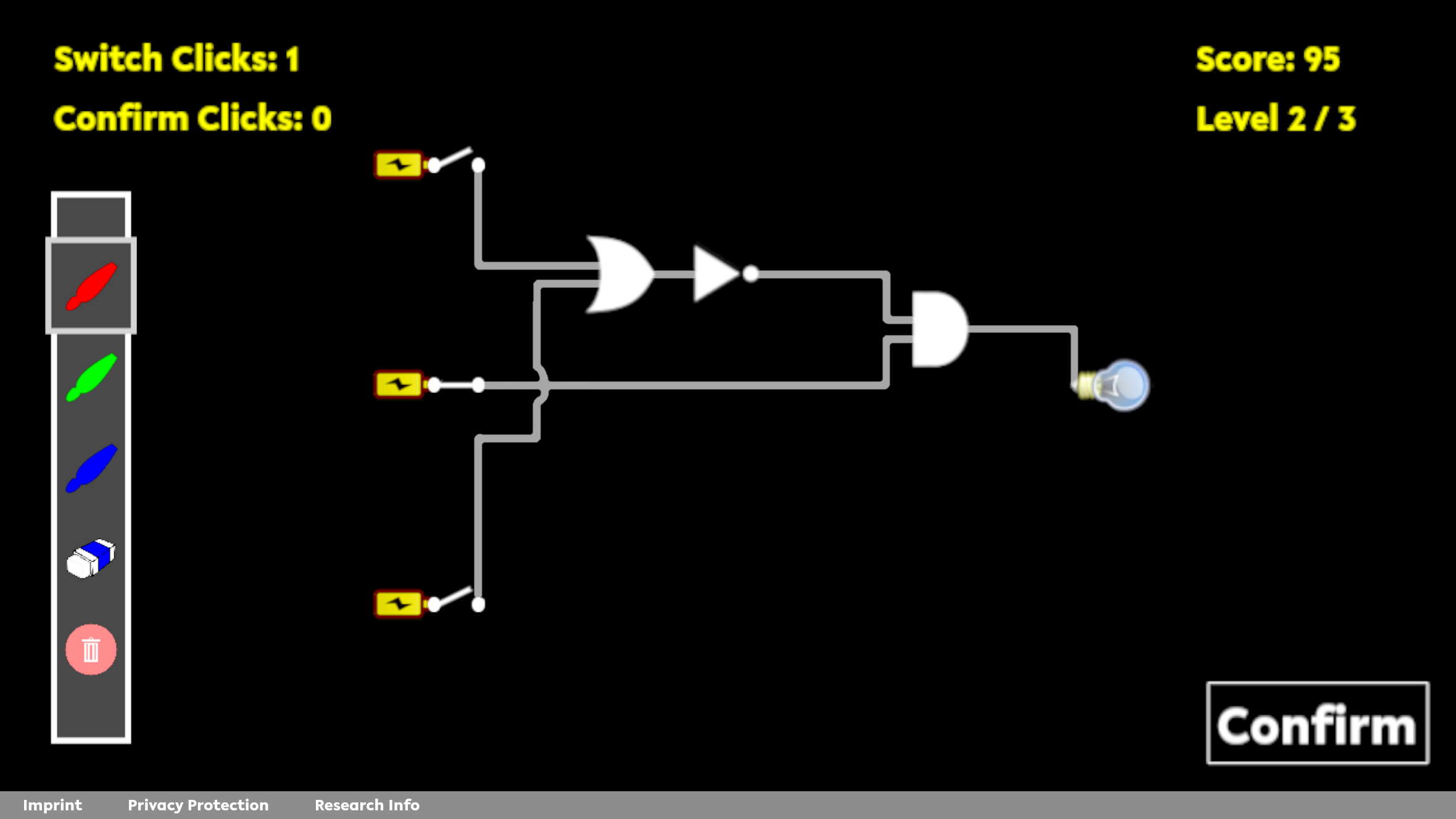}
    \caption{The interface for each task of \reversim consists of a Boolean circuit diagram with three inputs and at least one output. The participant interacts with the circuit by opening and closing the switches on the left. The example level shown here consists of three switches, three gates (AND, OR, and NOT) and one lamp as an output. Annotations can be drawn onto the circuit using the drawing tools on the very left. At the top, the participant's progress statistics are displayed.}
    \Description[screenshot of the user interface]{On a black background, the user interface displays the Boolean circuit for the present task in the center of the screen. From left to right, the circuit consists of three pairs of batteries and gates stacked vertically, three Boolean gate symbols (OR, NOT, AND) as defined in IEEE Std 91/91a-1991, and a single light bulb as an output. The circuit elements are connected using wires displayed as grey lines. On the very left of the screen, a vertical toolbar for annotating the circuit shows the red, green, and blue paint brushes, an eraser, and a trash can. On the top left, the environment displays two counters for the number of switch and confirm clicks. On the top right, the current score is shown together with the number of the current task. In the bottom right corner, there is a button labeled "Confirm". At the very bottom, a footer contains three buttons labeled "Imprint", "Privacy Protection", and "Research Info".}
    \label{fig:simulation:basicelements:gameui}
\end{figure}

The Boolean circuit itself is implemented through the three basic types of combinational gates, namely \textit{AND}, \textit{OR}, and \textit{NOT}.
Current flows through the circuit according to the Boolean functionality of these three gates and the \textit{wires} connecting them.
\autoref{fig:simulation:basicelements:gameui} shows a trivial example task that makes use of all three gate types.

\paragraph{Design considerations.} 
With our choice of basic elements, we are able to represent any combinational circuit as a task.
Those circuits -- \eg, in the form of logic functions located between memory elements of a large-scale netlist -- are an essential target for real-world analyses~\cite{meade2016netlist}.
The interaction mechanics of \reversim aim at modeling the sense-making processes that occur when reverse engineering such combinational subcircuits in the real world. 
The subcircuits' inputs and outputs, \ie, memory cells or external connections of the \ac{IC}, correspond to the switches, lamps, and danger signs in our tasks.

Real-world \ac{IC} manufacturing processes tend to use additional compound gate types for improved power and space efficiency~\cite{rahman2011power}, however, their functionality can generally be decomposed into the three basic gate types available in \reversim.
Although memory elements such as flip-flops could also be implemented using those basic gates, we decided to exclude sequential logic from \reversim for the time being to lower the entry barrier for participants with little prior knowledge.

\subsection{Obfuscated Gates}
\label{subsection:obfuscated}
\begin{figure}[t]
    \newcommand{\gatescale}{0.06}
    \centering
    \footnotesize
    \begin{tikzpicture}[scale=0.7]
        \node[inner sep=0pt,rotate=-90] (orgate) at (0,0)
            {\includegraphics[width=\gatescale\textwidth]{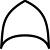}};
        \node[inner sep=0pt,rotate=-90] (inverter) at (4,0)
            {\includegraphics[width=\gatescale\textwidth]{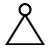}};
        \node[inner sep=0pt,rotate=-90] (camouflaged) at (0,2.5)
            {\includegraphics[width=0.07\textwidth]{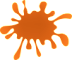}};
        \node[inner sep=0pt,rotate=-90] (andgate) at (4,2.5)
            {\includegraphics[width=\gatescale\textwidth]{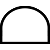}};
        \node[inner sep=0pt] (displayed) at (0,4)
            {displayed gate};
        \node[inner sep=0pt, align=center] (function) at (4,4.25)
            {example actual\\functionality};
        \node[draw=gray, thick, inner xsep=2.5em, inner ysep=0.8em, fit=(camouflaged) (andgate)] (groupCamouflaged) {};
        \node[fill=white,rotate=90] at (groupCamouflaged.west) {camouflaged};
        \node[draw=gray, thick, inner xsep=2.5em, inner ysep=0.8em, fit=(orgate) (inverter)] (groupCovert) {};
        \node[fill=white,rotate=90] at (groupCovert.west) {covert};
        \begin{scope}[on background layer]
            \draw[very thick]
                (-1, 2.2) -- ([yshift=-0.3cm]camouflaged.center)
                (-1, 2.8) -- ([yshift=0.3cm]camouflaged.center)
                (1, 2.5) -- (camouflaged.center);
            \draw[very thick]
                (-1, -0.3) -- ([yshift=-0.3cm]orgate.center)
                (-1, 0.3) -- ([yshift=0.3cm]orgate.center)
                (1, 0) -- (orgate.center);
            \draw[very thick]
                (3, 2.2) -- ([yshift=-0.3cm]andgate.center)
                (3, 2.8) -- ([yshift=0.3cm]andgate.center)
                (5, 2.5) -- (andgate.center);
            \draw[very thick]
                (3, 0.3) -| (3.25, 0) -- (inverter.center)
                (5, 0) -- (inverter.center);
            \draw[-{Rays[]}, very thick, dashed]
                (3, -0.3) -- ([xshift=-0.55cm,yshift=-0.3cm]inverter.center);
        \end{scope}
    \end{tikzpicture}
    \caption{The visualization of a camouflaged and a covert gate (left) and an example of their hidden functionality (right).}
    \Description[symbols for two types of obfuscated gates and examples for their true functionality]{The figure is split into two boxes. The upper one is labeled "camouflaged", while the lower one is labeled "covert". On the left side, each box shows the visual representation of the respective type of obfuscated gate. On the right, a logic gate symbol provides an example for the actual functionality of the respective obfuscated gate. In the case of the camouflaged gate, the obfuscated gate symbol is obscured by a large orange ink blot. Only the two input wires on the left and the output wire on the right can be seen. The actual functionality is shown by a symbol for an AND gate. For the covert gate, the obfuscated gate symbol is a two-input OR gate. The actual functionality is given by an inverter symbol where only the first input wire is connected, while the second input wire is not.}
    \label{fig:obfuscated:gatetypes}
\end{figure}

In addition to the three basic gates, we implemented two types of obfuscated gates in \reversim.
Both are derived from circuit obfuscation techniques previously proposed in the literature~\cite{cocchi2014circuit,shakya2019covert} and share the common goal of making their functionality difficult to determine through chip-level reverse engineering~\cite{becker2017hardware}.
\begin{itemize}
    \item Camouflaged gates~\cite{cocchi2014circuit} are a special set of gates designed for use in \acp{ASIC}. They implement the essential logic gates introduced in \autoref{subsection:basicelements} with the special property that all camouflaged gates look strikingly similar under microscopy. However, their appearance is clearly different from standard Boolean logic gates. While this makes them easy to identify as camouflaged gates, uncovering their individual Boolean functionality is hard. In \reversim, we represent such gates as an ink blot, drawing participants' attention to the fact that the gate is obfuscated, but not revealing its actual functionality. 
    \item Covert gates~\cite{shakya2019covert} have the additional property that they are \textit{not} easily identifiable as being obfuscated. Hence, a gate may appear as an \textit{OR} gate to the problem solver, but actually be an inverter, where only one of the two inputs is effective while the second is ignored. In \reversim, we represent such gates with the symbol of the gate that they pretend to implement.
\end{itemize}
\autoref{fig:obfuscated:gatetypes} shows how the camouflaged and covert gates are visually represented in a task and provides examples of the actual functionality that such gates may implement.

\paragraph{Design considerations.}
Obfuscation primitives such as camouflaged or covert gates are used in real-world designs to impede reverse engineering, as indicated by several existing patents~\cite{kuenemund2019semiconductor,chow2020secure,wang2020method}.
Including obfuscated gates in \reversim will allow researchers to explore approaches employed by analysts to reverse engineer individual obfuscation components, providing initial insight into the effective use of hardware obfuscation.

\subsection{Interaction Mechanics}
\label{subsection:mechanics}
The objective of each task is to choose the correct setting of each switch to control the respective outputs.
Once participants think they have set the correct switch positions, they can click the \textit{Confirm} button at the bottom right.
Participants are explicitly encouraged to only submit their solution once certain that it is correct, rather than trying all possible switch positions.
This is reinforced by a score in each task that gradually decreases with each incorrect submission, as well as a gradually increasing time delay before a new attempt can be made.
The effects of the current switch positions on the outputs are visualized as shown in \autoref{fig:simulation:mechanics:outputs} only after clicking the Confirm button.
During the tutorial and qualification phases only, current-carrying wires are additionally highlighted in yellow, making mistakes easier to spot.
If the solution was correct, the participant will move to the next task by clicking the \textit{Next} button.
If the solution was incorrect, the resulting circuit outputs are shown to the participant and they can start a new attempt by modifying their solution.
Should a participant submit multiple incorrect solutions for a task, or reach a specified timeout, they are offered the option to skip to the next task.

\paragraph{Design considerations.}
Determining (intermediate) values of individual wires is an essential subprocess in netlist reverse engineering~\cite{subramanyan2014reverse}.
A central part of our interaction mechanics is not allowing brute force and discouraging such trial-and-error strategies by imposing a score penalty and a time delay.
In real-world \ac{HRE}, using (automated) brute force may be a valid strategy to overcome some sufficiently small problems.
However, an analyst requires the cognitive capabilities to develop such custom automation and apply it to a particular netlist~\cite{wiesen2021anatomy}.
Disallowing brute force and enforcing manual interactions in \reversim may allow us to observe these cognitive aspects of \ac{HRE} problem solving.

Furthermore, it is often beneficial to analyze and validate the behavior of a netlist or part thereof using dynamic circuit analysis techniques without the need for costly experiments on a physical device~\cite{kimura2020from}.
Displaying the output state of the circuit even when an incorrect solution is submitted is therefore closely related to the dynamic analysis methods used in real-world \ac{HRE}.

The above mechanisms have been tuned through extensive piloting to identify sensible defaults.
However, each can be configured in detail to meet the exact requirements of various study objectives, environments, and participant groups.

\subsection{Task Design}
\label{subsection:leveldesign}
The \reversim software is accompanied by a library of tasks for use in the experiment phase, for which we provide details in \autoref{app:section:leveldetails}.
Each task consists of the basic elements introduced in \autoref{subsection:basicelements} in a circuit optimized to fit on a single screen.
We provide five sets of tasks, which include three different complexities -- low, medium, and high -- and two task types containing obfuscated gates.
\autoref{app:section:complexity} contains an example task for each of the different complexities.

Low complexity tasks have only one output, medium complexity tasks have two, and high complexity tasks have three outputs.
As the number of outputs increases, so does the number of gates in the circuit.
Each task contains three switches, regardless of complexity.
To ensure that outputs cannot be trivially determined and to introduce an additional measure of complexity, we also decided to require a minimum nonlinearity\footnote{The nonlinearity of a Boolean function is defined as the minimum Hamming distance to any linear or affine function~\cite{Sarkar00Nonlinearity}. The greater the distance, the more non-linear the function.} between inputs and outputs.
The same is required between all outputs, ensuring that no output follows trivially from another.
Tasks containing obfuscated gates are based on medium-complexity tasks, where a single gate has been replaced by a camouflaged or a covert gate.
\reversim also features a graphical task editor that allows researchers to extend and customize the task library.

\paragraph{Design considerations.}
Our tasks correspond to small modules of a larger netlist and have, for example, similarities to the transition logic of \acp{FSM}~\cite{fyrbiak2018difficulty} or to obfuscated \textit{micronetlists}~\cite{chow2020secure}.
However, the number of netlist components -- inputs, outputs, and gates -- is limited in \reversim, where the entire netlist is displayed on the screen in a fixed size.
Together with eliminating the need to navigate between modules, we are able to reduce the level of abstraction to a degree that is suitable for \ac{HRE} non-experts.

\subsection{Drawing Tools}
\label{subsection:drawingtools}
The participants can annotate each task with a few simple drawing tools provided on the left-hand side using their mouse.
This includes pens in three colors, as well as an eraser and the possibility to delete all previous annotations.
We chose colors with high and diverging contrasts to make them accessible.

\paragraph{Design considerations.}
Annotating a netlist has long been a central component of the sense-making process involved in \ac{HRE} and is therefore useful to observe.
Even before the advent of purpose-built reverse engineering software with annotation features, reverse engineers printed microscope photographs of \acp{IC} and then manually traced wires and gates~\cite{torrance2009state} with a pen.
Modern reverse engineering software provides advanced features for grouping, naming, or visually highlighting gates~\cite{fyrbiak2019hal,wallat2019highway}.
We opted for a simple drawing tool to avoid introducing additional complexity into the user interface. 
This way, \reversim remains accessible to participants who are not familiar with advanced netlist annotation features.

\subsection{Interactive Tutorial}
\label{subsection:tutorial}
To ensure that \reversim is accessible to participants with little prior knowledge of digital circuits, an interactive tutorial introduces all relevant elements and the objective of the tasks to the participants. 
For each circuit element from \autoref{subsection:basicelements}, the tutorial provides a textual and visual description.
It then encourages participants to individually try the elements out in minimal training circuits.
Here, all inputs to the respective gate are connected directly to a switch and battery, allowing the participant to directly manipulate each input.
Contrary to the general interaction mechanics, current-carrying wires are highlighted immediately, such that participants can observe the output behavior of the respective gate in real time.
This tutorial focuses specifically on explaining digital logic and circuit elements, as well as the user interface itself.
However, we deliberately chose to not present any \ac{HRE} concepts or specific solution strategies for the tasks that follow.

\paragraph{Design considerations.}
To successfully engage with an \ac{HRE} problem, a minimum knowledge of digital circuits and Boolean logic operators is required.
Therefore, the main objective of the tutorial is to ensure that each participant has a similar understanding of digital circuits and Boolean logic operators.
In particular, to be able to accurately observe strategies during \ac{HRE} problem solving, it is important to not introduce concrete approaches to solving the following tasks.
Doing so could bias participants towards the strategies included in the tutorial, and alter the decision-making process of which strategy to apply to a problem instance.

\subsection{Psychometric Test Integration}
\label{subsection:psychometric}
To test whether standardized cognitive tests can be integrated into \reversim and what they can reveal about the cognitive aspects of \ac{HRE}~(RQ3), we implemented a simple psychometric test very similar to the Trail Making Test~\cite{Bowie2006}.
The test we chose is a non-verbal intelligence test called ZVT~(German: ``\textit{Zahlen-Verbindungs-Test}''; number-connection test)~\cite{oswald2016zvt}, which reliably and validly measures (cognitive) processing speed, a component of intelligence~\cite{Vernon1993}.
In the original paper-and-pen test, participants use a pen to connect numbers from 1 to 90 on four standardized paper matrices as quickly as possible.
The implementation in \reversim includes an initial instruction page, two example matrices, and the four actual test matrices.
Participants used their mouse instead of a pen to click the numbers in order.
The two example matrices with numbers from 1 to 20 allowed participants to familiarize themselves with the task and interface.
Between the sample trials, participants were instructed to sit comfortably and click the numbers as quickly as possible.
To prevent the participant from mentally going through the entire number sequence and thus gaining a speed advantage, in all matrices only the numbers one to three are displayed before the first mouse click.

\paragraph{Design considerations.}
Cognitive processing speed is a cognitive factor that affects reasoning and the acquisition of new information~\cite{weiss2010theoretical} and thus likely influences \ac{HRE} performance, but the concrete interactions remain poorly understood.
Hence, it is a compelling focus for study as part of a broader effort to explore the diverse human aspects critical to \ac{HRE} success.
We opted for the number-connection test as it can be performed without experimenter intervention and without requiring audio or video capture, thus ensuring wide applicability and scalability.
When implementing the number-connection test in \reversim we carefully transferred all instructions from the test manual. 
As participants can easily track their progress on a pen-and-paper matrix, we made sure that, in our digital version, all correctly clicked numbers are clearly visually identifiable.
To make performance comparable across participants, we require that participants use a mouse when working on the number-connection test.
Thus, we avoid any performance influence resulting from the use of different input devices.

\section{Interview Study}
\label{section:interview}
To assess what aspects of real-world \ac{HRE} processes the \reversim environment models~(RQ1), we conducted semi-structured interviews with 14 researchers and professionals in hardware and netlist reverse engineering. 

\subsection{Methods}
\label{section:interview:methods}
After a brief review of ethical considerations, we describe the study participants, outline the interview procedures, and present our data analysis method. 

\subsubsection{Ethical Considerations \& Data Protection}
\label{section:interview:ethics}
Our institution's ethics committee and data protection officer approved the interview study.
All interviewees participated voluntarily, provided informed consent that included details of study procedures and data handling, and were free to withdraw from the study at any time.
Participants received the equivalent of \$48 in monetary compensation for time spent on study-related tasks.

\subsubsection{Participants}
\label{section:interview:methods:participants}
Our study involved 14 professionals and researchers in the field of \ac{HRE}. 
We reached out to participants in person at relevant hardware security conferences and via email.
Participants were pre-selected based on the following criteria: (a)~they had published significant work in the field, (b)~they were involved in publicly funded \ac{HRE} research projects, or (c)~they were engineers at companies specializing in \ac{HRE}.
Notably, none of the participants had been part of the continuous feedback loop during \reversim's initial development.

Our participants worked in academia~(8 participants from 3 institutions), federal agencies~(2 participants from 2 institutions), or international companies~(4 participants from 3 institutions) and were based in four different countries (France, Germany, Singapore, USA).
Twelve of the participants self-identified as male, two participants as female, and all had a university degree.
All participants self-rated their general \ac{HRE} expertise on a scale of 1~(novice) to 5~(expert) with a mean of $M = 4.0$ ($SD = .73$).
Overall, they had $M = 4.3$ years ($SD = 1.69$) of experience with \ac{HRE}. 
Half of the participants reported that they usually spend 20-30 hours a week or more for \ac{HRE} tasks.
In addition, all participants self-rated their prior theoretical \ac{HRE} knowledge (\eg, in Boolean algebra, \acp{IC}, or hardware obfuscation) and practical \ac{HRE} skills (\eg, analysis of data flows in netlists, or dynamic analysis of netlists) on a 5-point Likert scale ranging from 1~(very low) to 5~(very high).
We calculated two scales for self-ratings (see \autoref{app:section:interviewsubitems} for subitems).
The theoretical-knowledge scale had 10 items, and the practical-skill scale had 11 items.
Both scales had acceptably high Cronbach's alphas ($\alpha = .79$ for theoretical knowledge and $\alpha = .87$ for practical skills)~\cite{cronbach1951coefficient,tavakol2011making}.
Overall, the 14 participants self-rated their prior theoretical knowledge with a mean of $M = 3.9$ ($SD = .58$), and their practical skills in \ac{HRE} with a mean of $M = 3.7$ ($SD = .78$). 
In \autoref{appendix:experts}, we report detailed demographic and expertise data at the individual participant level.

\subsubsection{Study Procedure}
Researchers from hardware security and psychology collaborated to create and revise a semi-structured interview guide.
All interviews were conducted remotely by two researchers and lasted an average of 75 minutes.
The first interviewer had a background in cognitive science and experience in semi-structured interviews, the second in \ac{HRE}.

After being introduced to the procedure and objective of the study, \ie, the evaluation of \reversim by domain experts, the interviewees explored all main components of the \ac{HRE} environment (\ie, the interactive tutorial, the qualification phase, and the experiment phase). 
The experiment phase consisted of a total of five tasks designed to provide the participants with a comprehensive overview of \reversim's capabilities: One task each with low, medium and high complexity and two tasks containing obfuscated gates (see \autoref{subsection:obfuscated}). 

The final interview guide consisted of three question blocks focused on answering RQ1.
At the beginning of the interview, participants were asked about their first impressions of \reversim~(corresponding to Q1 in the interview guide).
In the main question block, the interviewers asked participants to compare the \reversim tasks with netlists they encounter in their professional work and how their applied strategies would compare~(Q3, Q4, and Q5).
The subquestions in Q3 and Q4 differed slightly depending on whether the participants had practical experience with netlist reverse engineering, which was asked in Q2.
The last block was about suggestions for improvements or future research with \reversim~(Q6). 

After the interviews were completed, participants answered an online questionnaire on demographics and prior knowledge as presented in \autoref{section:interview:methods:participants}.
Both the interview guide and questionnaire can be accessed online via \autoref{appendix:materials}.

\subsubsection{Data Analysis}
\label{section:interviewanalysis}
We transcribed the interviews verbatim and analyzed the transcripts based on the concept of qualitative content analysis~\cite{mayring2019qualitative} with two coders -- the interviewers -- in a two-stage process. 
In the first step, Coder~1 coded eight interviews to create an initial codebook, and the second coder familiarized themselves with the transcripts.
In collaboration, both coders iteratively created a final code book (see \autoref{appendix:detailedcodebook}) by defining code descriptions and rules as well as developing code categories for each question block.
In the second step, Coder 2 coded all 14 interviews, while Coder 1 revised their coding of the first eight and coded the remaining six interviews according to the final code book.
Finally, both coders compared their coding and discussed all deviations until consensus was reached.

\subsection{Results}
\label{section:interview:results}
Below we describe the code categories within the three question blocks in detail and support key statements with verbatim quotes from our participants. 

\subsubsection{Block 1: General Impression}
The first question block contains two core categories: \textit{Positive Feedback}, and \textit{Constructive Criticism}. 

When asked for their first impression, all participants gave \textit{Positive Feedback} on \reversim, stating that they enjoyed solving the tasks and liked the design of the environment (\eg, they found the drawing tools helpful). 
Furthermore, they mentioned that the interaction mechanics and objectives were clear and that \reversim was thoroughly implemented. 
Some interviewees reported that the abstraction in \reversim was realized very successfully (\eg, the implementation of the danger sign and lamp symbol as outputs of the circuits). 
Many participants explicitly mentioned that the interactive tutorial was didactically well-structured and indicated that they felt well guided by the tutorial. 

Some participants also raised \textit{Constructive Criticism}. 
A few mentioned that an additional explanation of the gates and their functions should be available on an ongoing basis. 
In their opinion, this type of backup would be very helpful especially for non-expert participants, so that they could look up the functions, \eg, of an AND gate. 
Furthermore, some participants suggested revising the tasks containing obfuscated elements. 
Most of the participants solved these obfuscated tasks without having to consider or identify the obfuscated elements. 
Accordingly, these participants suggested that obfuscated gates be placed in the critical path of analysis. 
Lastly, some of the participants suggested improvements in the handling of user input, particularly in the use of the drawing tools. 

\subsubsection{Block 2: Comparing \reversim with Reality}
In the second question block, five main categories resulted from our analysis: \textit{Elements Recognized from Real-World Netlist Reverse Engineering}, \textit{Real-World Elements Missing in \reversim}, \textit{Real-World Approaches Covered by \reversim}, \textit{Real-World Approaches Missing in \reversim}, and \textit{Evaluation of Interaction Mechanics}. 

Regarding the identification of \textit{Elements Recognized from Real-World Netlist Reverse Engineering}, all participants recognized components in the environment that they knew from analyzing real-world netlists (\eg, logic gates such as AND or OR gates; input-output relations). 
One of the participants summarized this point with the following words: ``The concept of making the light bulb glow is the same as making a bit in a netlist one; and setting the bit to zero in a netlist is the same like for the danger symbol in a task. 
That actually occurs in netlists, because you often have an output behavior, where you just want to know `Okay, what do I need now as input configuration, so that a certain output is zero and a certain output is one?'.'' 

Several interviewees also mentioned \textit{Real-World Elements Missing in \reversim}. 
A few participants said that they usually have to analyze netlists that also contain further logic gates such as NANDs, NORs, or XORs, as well as sequential logic elements (\eg, flip flops) and that these did not appear in the tasks. 
Some participants mentioned that the complexity of the tasks was only partly comparable to the complexity of real netlists with up to millions of gates. 
However, those participants also reported that they typically use semi-automated tools or write scripts to reduce complexity in real netlists. 
Finally, a few participants added that they had never seen obfuscated elements in real-world netlists.
However, they also indicated that solving the obfuscated tasks reminded them of dealing with erroneous netlists. 

We identified several answers mentioning \textit{Real-World Approaches Covered by \reversim}. 
Most participants rediscovered problem-\linebreak{}solving processes they usually apply in real-world netlist analysis when solving tasks in \reversim.
For example, interviewees indicated that they proceeded backwards from the outputs to the inputs.
Such output-driven approaches and back justifications are -- according to the participants -- common practice in \ac{HRE}.
Some participants hypothesized about how a particular output depends on specific inputs in \reversim and then annotated them with the drawing tools. 
According to the experts, this procedure is comparable to the procedure in real netlists, where hypotheses about input-output relations are formulated and tested.  
One participant stated: ``I wasn't quite sure whether the switch had to be set to one or zero or whether there was another possibility. And then I just set an additional input to zero or one and then I looked if that can work or not. And if it doesn't work out, then it must be the other one.''
Another participant noted that they usually apply annotations comparable to the annotation processes in \reversim: ``So in \ac{HRE} practice, I would just do something like that by annotating any signals on it. But do the whole thing in a framework and not on the screen. But the concept is the same\dots it's just a bit more work with the mouse if you don't have a tablet and a pen.''

Some participants reported \textit{Real-World Approaches Missing in \reversim}.
A few indicated they would regularly use forward analysis or brute-force approaches in real netlists. 
However, applying a brute-force approach in \reversim was discouraged by deducting points. 
In addition, some interviewees indicated that they would develop semi-automated approaches, especially for increasingly complex real-world netlists. 
Nevertheless, participants added that our tasks have practical relevance, explaining that the tasks represent the manual analysis in the \ac{HRE} process after the complexity of the circuit had been reduced by semi-automated steps and scripts.

Participants also provided an \textit{Evaluation of Interaction Mechanics}. 
Most participants stated that the interaction mechanics did not force them to apply unrealistic steps that they would not apply in real-world netlist analysis. 
However, the fact that brute forcing was penalized was viewed ambivalently, as some participants would have liked to be free in their approaches without deduction of points. 
On the contrary, some participants drew a comparison and would consider brute force inefficient in real netlists, as summarized by this participant: 
``If I transfer the whole thing to real problems, \ie, large netlists, then brute forcing -- with a correspondingly large input space -- would not be possible.''

\subsubsection{Block 3: Future Research and Features}
The third block includes statements about possible future research and features for \reversim and consists of two main topics: \textit{Add Further Gates and Netlist Components} and \textit{Additional Objectives for Tasks}.

Some participants mentioned that future tasks of \reversim could include further combinational gates (\eg, NAND, XOR), sequential gates (\eg, flip flops), or high-level components such as adders. 
However, most participants concurred that incorporating additional components could be also more difficult for participants with little prior domain-specific knowledge. 
Further, a few participants expressed the idea to incorporate \textit{Additional Objectives for Tasks}. 
For example, different netlist modules could be presented, from which the one implementing a certain functionality should be selected.
One participant described this idea as follows: ``Maybe you give like a list of possible modules like multiplexer, addition, \elide some basic function. And then you ask the user: Which one applies to your circuit?''

\subsection{Discussion of RQ1: \reversim Models Key \acs{HRE} Aspects with Room for Expansion}
\label{section:interview:discussion}
Our interview study indicates that \reversim effectively models several key aspects of real-world netlist reverse engineering, although with some important caveats.
The positive feedback from experts suggests that \reversim is both engaging and didactically sound, successfully mirroring certain problem-solving processes and strategies that are fundamental in \ac{HRE}.
In particular, the challenge of analyzing circuits in \reversim was found to be comparable to real-world tasks, and the interaction mechanics, including the use of drawing tools, were recognized as intuitive and helpful for facilitating the problem-solving process.
However, participants also noted limitations in \reversim's ability to fully capture the complexities and nuances of actual netlist analysis.
The simplified nature of the tasks, while effective for modeling specific subprocesses of \ac{HRE}, does not entirely reflect the complexity and scale of real-world netlists.
According to our participants, these subprocesses are indeed relevant in practice, though they are more representative of a focused, manual analysis stage where reverse engineers have already reduced the complexity in advance (see our design considerations in \autoref{subsection:leveldesign}).
Moreover, the deliberate discouragement of brute-force approaches in \reversim, intended to promote thoughtful analysis, contrasts with real-world practices where brute-force methods are sometimes employed, particularly in preliminary analysis stages.
This design choice was intentional, as outlined in \autoref{subsection:mechanics}, to encourage participants to carefully consider the origins of individual outputs, thus mirroring critical thought processes in \ac{HRE}.
Looking ahead, \reversim could benefit from incorporating additional components and objectives to more comprehensively capture the broader challenges encountered in real-world netlist reverse engineering, such as introducing more complex logic gates, sequential elements, and higher-level components.
While we plan to incorporate these more extensive feature requests in future open-source development, we have promptly addressed several smaller issues identified by participants, including revising tasks involving obfuscated gates and improving the usability of the drawing tools.

\section{User Study}
\label{section:pilot}
To assess the minimum requirements for effective participation in \reversim~(RQ2a), to evaluate whether participants' \ac{HRE} performance can be differentiated based on the tasks included in \reversim~(RQ2b), and to examine the integration of standardized cognitive tests~(RQ3a) and what insights they provide about the human aspects of \ac{HRE}~(RQ3b), we conducted a user study with 109 participants.

\subsection{Methods}
\label{section:pilot:methods}
After outlining the ethical considerations, we provide information about the study participants and procedures as well as the measures and variables collected for this study.

\subsubsection{Ethical Considerations \& Data Protection}
\label{section:pilot:ethics}
The user study was approved by the ethics committee and data protection officer of our institution.
All participants took part voluntarily, provided informed consent that included details of study procedures and data handling, and were free to withdraw from the study at any time.
To link participants' answers in the questionnaires to their log files, we assigned randomly generated pseudonyms.
We also ensured encrypted communication between the clients and the server and stored the study-related materials on an internal server to which only the researchers involved in the study had access.

\subsubsection{Participants}
\label{section:pilot:methods:participants}
We recruited a total of 131 participants from the US and UK via Prolific\footnote{Prolific is an online platform that connects researchers with a diverse pool of participants for studies; see \url{https://www.prolific.com/}.}, divided into a pilot sample of 20 and a main sample of 111 participants, of which we had to exclude two due to technical issues.
The eligibility criteria for participation in the study were a minimum age of 18 years, fluency in English, and a minimum level of education equivalent to a university entrance qualification.
Participants were further required to use a desktop or laptop PC with a mouse.
Out of 109 participants, 10 dropped out while working with \reversim, resulting in 99 complete data sets (see \autoref{fig::simsetting:flowchart});
Participants received \textsterling 22.50\footnote{Prolific always bills in British pounds.} as compensation for completing the study and spent a median of $69.5$ minutes doing so.

Of our final sample ($n=99$), 70 participants self-identified as male and 29 as female.
Participants were between 18 and 72 years old, with a mean age of $M = 38$ ($SD = 12.3$).
72 participants had a university degree, eight had a professional degree, 18 had a high school degree or equivalent, and one preferred not to disclose.

\subsubsection{Study Procedure}
\label{subsubsection:simsetting}

\begin{figure*}[ht!]
    \footnotesize
    \centering
    \begin{tikzpicture}[node distance=0.2cm and 0.7cm]
        \node[draw, align=center, thick] (presurvey) {Pre Survey\\($n=109$)};
        \node[draw, align=center, right = 4em of presurvey] (zvt) {Number\\Connection Test};
        \node[draw, align=center, right = of zvt] (tutorial) {Interactive\\Tutorial};
        \node[draw, align=center, right = of tutorial] (quali) {Qualification\\Phase};
        \node[draw, align=center, right = of quali] (competition) {Experiment\\with 12 Tasks};
        \node[draw, align=center, right = of competition] (end) {End};
        \node[draw, align=center, right = 4em of end] (postsurvey) {Post Survey\\($n=99$)};

        \draw[->] (quali) edge [dashed, bend right=50] node [midway, above, align=center] (requalifylabel) {repeat tutorial voluntarily\\or if not qualified ($84$)} (tutorial);
        \draw[->] (quali) edge [dashed, bend left=30] node [pos=.78, above, align=center] (to1label) {timeout ($2$)} (end);
        \draw[->] (competition) edge [dashed, bend right=40] node [midway, below, align=center] (to2label) {timeout ($22$)} (end);

        \node[draw=gray, thick, inner xsep=1.5em, inner ysep=0.75em, fit=(zvt) (tutorial) (quali) (competition) (end) (requalifylabel) (to1label) (to2label)] (ingame) {};
        \node[fill=white, align=center] at (ingame.north) {\reversim Environment};
        
        \node[align=center, below = 4.5em of quali] (dropout) {dropout (10)};
        
        \draw[->](presurvey) edge node[pos=.25, above] {109} (zvt)
            (zvt) edge node[midway, above] {108} (tutorial)
            (tutorial) edge node[midway, above] {108} (quali)
            (quali) edge node[midway, above] {98} (competition)
            (competition) edge node[midway, above] {75} (end)
            (end) edge node[pos=.70, above] {99} (postsurvey);

        \draw[->,densely dotted] (zvt) |- node[pos=.1, left] {1} (dropout);
        \draw[->,densely dotted] (quali) edge node[pos=.25, left] {8} (dropout);
        \draw[->,densely dotted] (competition) |- node[pos=.1, left] {1} (dropout);
    \end{tikzpicture}
    \caption{Overview of the flow of our user study. Excluding two invalid datasets, 109 participants started the study. 84 revisited the tutorial at least once. Ten participants dropped out during the different phases of the study, particularly during the qualification phase, resulting in a total of 99 valid and complete datasets. 24 participants did not finish within the 75-minute time limit before proceeding to the post survey.}
    \Description[Flow of the user study]{The flow chart consists of 8 states of which 5 are inside the ReverSim environment. All participants start in "Pre Survey" and end up in either the "Post Survey" or "dropout" state. All 109 participants follow the main path inside the ReverSim environment to "Number Connection Test", 108 enter "Interactive Tutorial", all 108 move along to "Qualification Phase", 98 continue to "Experiment with 12 Tasks", 75 reach "End" and 99 continue to "Post Survey". Two participants from "Qualification Phase" and 22 from "Experiment with 12 Tasks" also joined the "End" state due to a timeout. There was one dropout in "Number Connection Test", 8 in "Qualification Phase" and 1 in "Experiment with 12 Tasks". 84 Participants were thrown back from "Qualification Phase" to "Interactive Tutorial" voluntarily or because they did not qualify for the next state.}
    \label{fig::simsetting:flowchart}
\end{figure*}

Participants proceeded through the study as shown in \autoref{fig::simsetting:flowchart}.
After voluntarily consenting to participate in the study, all participants answered a pre-study questionnaire. 
First, the questionnaire screened if they met the eligibility criteria.
If so, we asked them about further demographics, including their academic and professional education, experiences in computer science and the \ac{IC} industry.
Participants then self-rated their prior knowledge in 16 relevant areas (\eg, Boolean algebra, logic gates, reverse engineering; see \autoref{appendix:playabilitysubitems} for a complete list) on a scale of $0$~(``none'') to $5$~(``very high'').
Afterwards, they completed the four phases of \reversim (see \autoref{fig:simulation:flow:diagram}), starting with the number-connection test~\cite{oswald2016zvt}.
During the experiment phase, participants engaged in 12 \ac{HRE} tasks of increasing difficulty, which we selected based on mean and variance of solution times and participants' feedback gathered during development:
Two low complexity tasks~(Group~A), four medium complexity tasks~(Group~B), four high complexity tasks~(Group~C), and two obfuscated medium complexity tasks~(Group~D) (see \autoref{subsection:leveldesign} and \autoref{app:section:leveldetails} for detailed information about each task).
We randomized the order of the tasks within each group to limit order bias; and for reasons of fair payment, the total time spent in \reversim was limited to 75 minutes.
After working with \reversim, we asked study participants to provide feedback on positive and negative aspects of the environment.
All questionnaires are available in the online appendix (see \autoref{appendix:materials}).

\subsubsection{Measures and Variables}
\label{section:pilot:methods:measures}
\paragraph{Prior Knowledge.}
To assess participants' domain-specific prior knowledge, we aggregated their responses to all 16 prior-knowledge topics into a single mean score.
This was possible due to their high internal consistency, as evidenced by the extremely high Cronbach's $\alpha$~\cite{cronbach1951coefficient} of $0.96$, indicating that these items measure the same underlying construct~\cite{tavakol2011making}.

\paragraph{\ac{HRE} Performance Variables.}
To measure participant performance, we recorded log files during the experiment phase.
From this data, we calculated the following set of variables for each of the 12 tasks:

The \textit{Task Solved} variable captures whether a participant has solved a task.
Tasks may remain unsolved if the participant skips the task or if the study times out after 75 minutes.
To avoid stalling participants, tasks become skippable after four unsuccessful attempts, or after exceeding a predetermined time within the task.
All time limits are listed in \autoref{app:section:leveldetails}.
To ensure data quality, we consider a task that has been solved by brute force, \ie, by rapidly trying arbitrary switch combinations without thinking, as not solved.
We identify a task as brute-forced if we observe an average attempt rate of more than one submission within 10 seconds.

For all tasks that were solved, we calculated the following two variables:
The variable \textit{Time in Task} measures the time until a participant submits a \textit{correct} solution.
The variable \textit{Number of Attempts} measures the number of attempts it takes the participant to produce this correct solution.

\paragraph{Cognitive Processing Speed.}
From the number-connection test (see \autoref{subsection:psychometric}), we determined the \textit{Time per Matrix} by measuring the seconds between clicking '1' and completing each of the four sequences.
We then assessed participants' \textit{Cognitive Processing Speed} by calculating the mean Time per Matrix for each participant.
To evaluate whether the integration of the number-connection test into \reversim was successful, we compared participants' performance with the norm values and corresponding \ac{IQ} values for the cognitive processing speed subcomponent of intelligence from the original test manual.
To investigate possible correlations between processing speed and task solution speed we picked one medium-complexity task.
We calculated three correlations, each with processing speed: time to solve the task on the first attempt, time to solve the task regardless of the number of attempts, and how many \ac{HRE} tasks a participant was able to solve in total.
We chose Pearson's correlation for solution speed and Spearman's rank correlation as well as Kendall's $\tau$ for \ac{HRE} tasks solved as the latter variable is an ordinal scale.
Positive correlations were assumed for the time variables (\ie, less time to complete the matrices relates to less time to solve a task) and a negative correlation for the amount of tasks solved (\ie, less time to complete the matrices relates to more solved tasks).

\subsection{Results}
\label{section:pilot:results}

This section presents a detailed analysis of participants' prior knowledge, engagement, and performance in \reversim. 
We further assess the integration of a standardized cognitive test by analyzing the number-connection test results and examining their relationship to participants' \ac{HRE} task performance.
%\pagebreak

\subsubsection{Participants' Prior Knowledge \& Experience}
\begin{figure}[ht]
    \centering
    \includegraphics[width=.95\columnwidth]{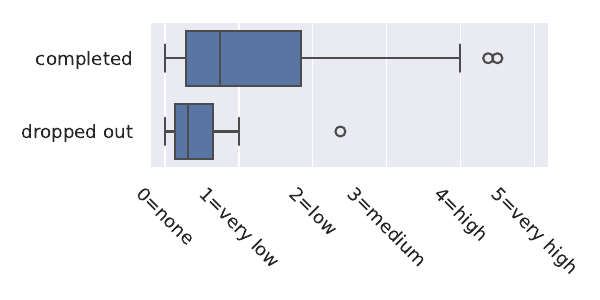}
    \caption{Distribution of prior-knowledge scores among our 109 participants. The 99 participants who completed the study had a median score of $0.75$, while the ten participants who dropped out had a median score of $0.31$. Nine of the ten dropouts had a prior knowledge score of $1$ or less.}
    \Description[Distribution of prior-knowledge among participants]{Box plots visualizing the prior knowledge scores for the group who completed the test and those who dropped out. The scale is divided into 6 steps ranging from 0 to 5: "none", "very low", "low", "medium", "high" and "very high". The whiskers for the completed group range from 0 to 4, with two outliers around 4.5. The whiskers for the dropped-out group range from 0 to 1, with one outlier at 3.5. The interquartile range goes from slightly above "none" to just below "low" for the participants who completed the test and from almost none to just above "very low" for those who dropped out.}
    \label{fig:pilot:methods:participants:pkscore}
\end{figure}

Our 109 participants (including dropouts) self-rated their prior knowledge with a mean of $M = 1.10$ ($SD = 1.08$, $median = 0.75$), corresponding to ``very low''.
The distribution of prior knowledge among our participants, broken down by those completing the study and those who dropped out, is shown in \autoref{fig:pilot:methods:participants:pkscore}.
42 participants reported practical experience in computing and nine reported practical experience with microchips.

\subsubsection{Participant Engagement \& Performance}

Of our 109 participants, 99 -- or 91\% -- completed the study and ten participants dropped out.
As shown in \autoref{fig::simsetting:flowchart}, eight participants dropped out during the qualification phase, one dropped out during the number-connection test, and one dropped out during the experiment phase.
Of the ten dropouts, only one participant had a background in computing, and none had experience with microchips.

\begin{figure}[ht]
    \centering
    \includegraphics[width=.95\columnwidth]{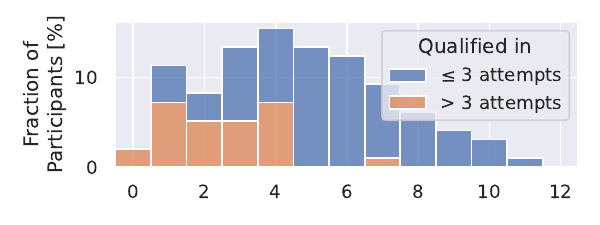}
    \caption{Number of \ac{HRE} tasks solved on the first attempt by fraction of participants, broken down into those requiring a maximum of three attempts to qualify and those requiring more than three attempts.}
    \Description[Number of HRE tasks solved on first attempt by participants who needed less than four or four and more qualification attempts]{The histogram shows the number of tasks on the x-axis and the fraction of participants on the y-axis. Each bar is split into a fraction of participants who qualified in less than four attempts and into participants who needed four or more attempts. Over the number of solved tasks on the first attempt, the fraction of participants that required three or fewer qualification attempts follows a normal distribution centered around five solved tasks. On the other hand, the participants requiring more than three attempts are almost evenly distributed between 1 and 4 solved tasks.}
    \label{fig:results:attempts_distribution}
\end{figure}

In the following, we only consider the 99 participants who completed the study, 24 of whom reached the time limit of 75 minutes, and in consequence missed a median of two tasks.
Participants solved a mean of eight (out of twelve) tasks with any number of attempts, and a mean of $M=4.5$ tasks on first attempt, which is a very strict measure of success.
\autoref{fig:results:attempts_distribution} shows the number of tasks solved on first attempt, which ranges from 0 to 11, by fraction of participants.
Remarkably, the 27 participants requiring more than three qualification attempts solved a mean of just $M=2.5$ tasks on first attempt, which is only marginally better than random guessing, while the 70 participants requiring three or less qualification attempts solved a mean of $5.5$ tasks.

\subsubsection{Participants' Feedback}
Below we report the feedback provided by all 99 non-dropout participants in the post-survey.
No participants reported accessibility issues when using \reversim.
On a scale of 1~(``fully disagree'') to 5~(``fully agree''),  participants agree~($M = 4.06$, $SD = 1.10$) that they enjoyed engaging with \reversim.
They also agree that they understood the interaction mechanics~($M = 4.36$, $SD = 0.81$) but were undecided whether the scoring was motivating ($M = 3.39$, $SD = 1.14$).
Participants appreciated the opportunity to repeat the tutorial ($M = 4.35$, $SD = 0.79$) and agreed that it was easy to understand ($M = 4.10$, $SD = 1.04$).
A total of 74 participants indicated having used the drawing tools, agreeing that they are useful ($M = 4.16$, $SD = 0.95$).

\subsubsection{Per-Task Performance}
For this analysis, we only consider the 73 participants who did not reach the 75-minute time limit and thus had sufficient time to attempt each task.
This intentional choice prioritizes the comparability of tasks, ensuring that any observed variations are due to task differences rather than ordering effects.

\begin{figure}[ht]
    \centering
    \includegraphics[width=.95\columnwidth]{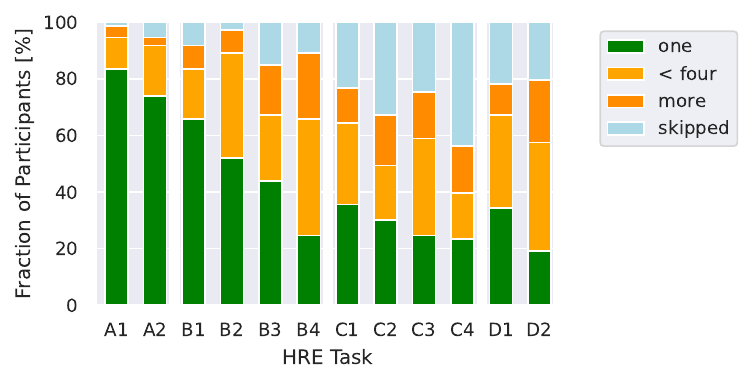}
    \caption{Fractions of participants who solved each task by the number of solution attempts. Participants were allowed to skip tasks after four unsuccessful attempts or after exceeding a predetermined time within the task (see~\autoref{app:section:leveldetails}).}
    \Description[Number of attempts for each task]{The bar chart consists of one bar for each task. The task name (A1-2, B1-3, C1-3 and D1-2) is on the x-axis and the percentage of participants on the y-axis. Each bar is divided into four sections: green for one attempt, yellow for less than 4 attempts, orange for more than four attempts and light blue for those who skipped the tasks. There is a decrease in the size of the green bars, and increase of light green bars. This ranges from 84\% needing only one attempt in the first task down to 19\% in the last task. The percentage of participants requiring multiple attempts increases until the fourth task and then remains relatively consistent.}
    \label{fig:results:attempts}
\end{figure}

\autoref{fig:results:attempts} shows the fractions of participants who solved each task, broken down by the number of attempts.
If we ignore the number of attempts, then Group~A tasks were solved by 99\% to 95\% of the participants, Group~B tasks by 85\% to 97\%, Group~C tasks by 56\% to 77\%, and Group~D tasks by 78\% to 79\%.
If we consider only successful solutions on the first attempt, the solution probabilities for Group~A tasks are 74\% to 84\%, for Group~B tasks 25\% to 66\%, for Group~C tasks 23\% to 36\% and for Group~D tasks 19\% to 34\%.

\begin{figure}[ht]
    \centering
    \includegraphics[width=.95\columnwidth]{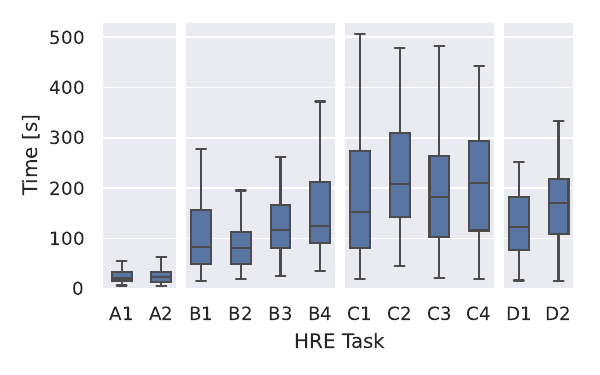}
    \caption{Time required by the participants to solve each \ac{HRE} task. Of the 73 participants, only the times of those participants who solved each individual task correctly were included, \ie, sample size varies between 72 participants for Task A1 and 28 participants for Task C3.}
    \Description[Time required to solve each task]{The box plot shows the time participants spent on each task, only taking those into account who solved the task correctly. The x-axis shows the task names and the y-axis shows the number of seconds required to solve the task. For tasks A1-A2 participants took under 100 seconds to solve the task with a mean well below 50 seconds. For tasks B1-3, participants took up to nearly 400 seconds, with the mean time lying between 80 and 120 seconds. Tasks C1-C3 took up to 600 seconds, and the mean times for each task were between 150 and 220 seconds. Finally, for tasks D1-D2 up to 340 seconds were required to complete the tasks with a mean between 120 and 180 seconds.}
    \label{fig:results:timing_subset}
\end{figure}

\autoref{fig:results:timing_subset} shows the distribution of times to correct solutions for each \ac{HRE} task.
Median times were $22$ to $23$ seconds for Group A tasks, $81$ to $124$ seconds for Group B tasks, $153$ to $210$ seconds for Group C tasks, and $123$ to $170$ seconds for Group D tasks.

%\pagebreak
\subsubsection{Number-Connection Test Results}
108 participants completed all four matrices of the number-connection test.
Participants solved Matrix~1 in a mean time of $M=66$ seconds (range: $39$ to $133$), the second in $M= 66$ seconds (range: $37$ to $137$), the third in $M= 72$ seconds (range: $42$ to $467$), and Matrix~4 in $M= 90$ seconds (range: $43$ to $611$).
In accordance with the test manual we excluded two participants who exceeded a total time of ten minutes working on the matrices.
All distributions show a positive skew, \ie, an above-average number of participants with short solution times, which is in line with expectations based on the test manual.
However, a Welch's ANOVA~\cite{Welch1951} ($F = 4.63, df = 3, p < .001$) and Bonferroni-corrected pairwise t-tests revealed a significant difference in means between Matrix 4 and all other matrices (each $p < .001$).
As the manual suggests excluding matrices when participants struggle with them -- albeit on an individual basis --  we extended this principle to the group level and excluded Matrix~4 from further calculations to enhance reliability and validity of our findings.
Using matrices 1-3, the mean processing speed subcomponent of our participants \acp{IQ} -- as estimated by the number-connection test manual -- is about $118$, which is about one standard deviation above average (\ie, $100$).

\subsubsection{Correlation between Cognitive Processing Speed and \texorpdfstring{\acs{HRE}}{HRE} Task Performance}
\begin{figure}[t]
    \centering
    \includegraphics[width=.95\columnwidth]{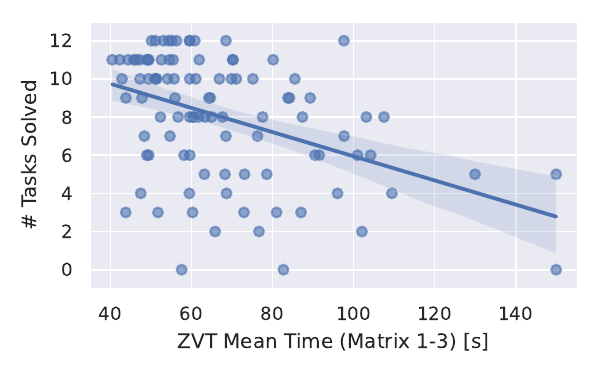}
    \caption{Mean time spent on the first three matrices of the number-connection test (x-axis) and the number of tasks solved with any number of attempts (y-axis). We show the regression line to provide an intuition of the overall trend.}
        \Description[number of tasks solved in correlation with mean time spent on the first three matrices of the number-connection test]{The x-axis of the scatter plot shows the time in seconds that novice participants needed to complete the number-connection test. It ranges from 40 to 150 seconds. The y-axis shows the number of tasks solved ranging from 0 to 12. There is a high density of points in the top-left corner, showing many participants solving eight or more tasks in fewer than 80 seconds. The density of the point cloud decreases with increasing time and decreasing solves, as only three participants required more than 120 seconds and two solving less than two tasks. The regression line shows a slight downward trend, meaning fewer tasks were solved by participants who took more time to complete the test. It ranges in a straight line from 10 tasks solved in 40 seconds to 3 tasks solved in 150 seconds.}
    \label{fig:results:zvt_corr}
\end{figure}

\begin{figure}[t]
    \centering
    \includegraphics[width=.95\columnwidth]{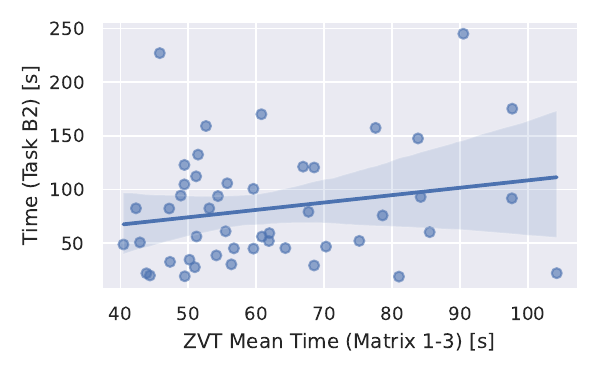}
    \caption{Mean time spent on the first three matrices of the number-connection test (x-axis) and \textit{Time in Task}~B2, limited to those participants solving the task on first attempt (y-axis). We show the regression line to provide an intuition of the overall trend.}
    \Description[time to solve task B2 first try in correlation with mean time spent on the first three matrices of the number-connection test]{Scatter plot showing the time in seconds that novice participants needed to complete the number-connection test on the x-axis. It ranges from 40 to 105 seconds. The y-axis shows the solution time for all players who have finished task B2 on the first attempt. It ranges from 0 to 250 seconds. There is a high density of points in the bottom-left corner. The density of the point cloud decreases with increasing ZVT Mean Time. The regression line shows a slight upward trend, meaning participants who were quick in the ZVT were also faster in solving level B2 on the first try. It ranges in a straight line across the whole ZVT Mean Time range starting at 70 seconds for task B2 and ending at 110 seconds.}
    \label{fig:results:zvt_corrtime}
\end{figure}

To investigate the relationship between participants' processing speed and their \ac{HRE} task performance in terms of solution speed and solution probability, we calculated three correlations (see \autoref{section:pilot:methods:measures}).
The correlation between processing speed and the total number of \ac{HRE} tasks solved by each participant, visualized in \autoref{fig:results:zvt_corr}, was assessed using Spearman's rank correlation ($\rho = -0.38$) and Kendall's $\tau$ ($\tau = -0.27$), both indicating a moderate correlation.
We then picked Task~B2 as an example, because of its high solution probability (see \autoref{fig:results:attempts}) and largely symmetric variance in solution time (see \autoref{fig:results:timing_subset}).
A Pearson correlation between processing speed and \textit{Time in Task}~B2 for participants solving the task on first attempt revealed a weak positive correlation ($r = .21 $), visualized in \autoref{fig:results:zvt_corrtime}.
A similar Pearson correlation for processing speed and \textit{Time in Task}~B2 regardless of the number of attempts also showed a weak positive correlation ($r = .11 $).


\subsection{Discussion of RQ2: \reversim is Open to Non-Experts and Differentiates Their \acs{HRE} Performance}
\label{section:pilot:discussionrq2}

\paragraph{RQ2a: Minimum Requirements for Participation}
Despite participants' very low prior knowledge in relevant areas, many were able to meaningfully engage with \reversim, as evidenced by a low dropout rate of less than 10\% and a high mean number of eight tasks solved per participant.
We did not find any significant influence of prior knowledge or experience on performance -- likely due to the uniformly low levels of prior knowledge -- yet observed an interesting pattern among dropouts: 
Participants with a prior knowledge score above $1$ or with experience in computing or microchips had a very low probability of dropping out. 
Future studies using \reversim could therefore pre-select participants accordingly.

Approximately three-quarters of the participants revisited the tutorial, and they consistently agreed that it was easy to understand.
This result suggests that while the tutorial effectively conveys the interaction mechanics, the qualification phase remains a challenging test of participants' ability to devise their own \ac{HRE} strategies.
Interestingly, even participants who required up to three qualification attempts were generally successful during the experiment phase, according to the strict metric of \textit{Task Solved on First Attempt}. 
However, those who needed more than three attempts to qualify were largely unable to engage effectively with \reversim.
This finding leads us to recommend early screening of such participants, while ensuring they are compensated proportionally for their participation.
By applying these screening criteria, the number of tasks solved on the first attempt is more evenly distributed across the available range, enhancing the expressiveness of the resulting data.

Approximately 22\% of participants reached the time limit during the experiment phase.
However, we find that they were generally still able to engage well with \reversim, as evidenced by a mean number of $6.5$ ($sd = 2.3$) correctly solved tasks prior to reaching the timeout.
Therefore, it will be worthwhile to extend the time limit in future studies, provided that some other mechanism to ensure fair compensation of participants' time is in place.

\paragraph{RQ2b: Differentiation of Participants' \ac{HRE} Performance Based on Tasks}
To evaluate whether participants' \ac{HRE} performance could be differentiated based on the tasks included in \reversim, we analyzed solution probability and solution time as key metrics.

We observed large differences across tasks for both metrics, underscoring the range of challenges presented by the different tasks.
Specifically, from Group~A to Group~D, we observed a strong decrease in solution probability, indicating that our task groups span a wide range of difficulty levels.
The percentage of participants who solved a task on first attempt emerged as a particularly informative metric, ranging from 84\% to 19\%. 
This range demonstrates that \reversim can differentiate participant performance across the entire difficulty spectrum without encountering ceiling or floor effects.
Within Group B, we make an interesting secondary observation: 
Even though all four tasks were designed based on our medium-complexity design criteria (\autoref{subsection:leveldesign}), solution probability on first attempt varied by as much as 41\% from task B1 to B4.
Hence, factors beyond the number of outputs and Boolean nonlinearity appear to influence task difficulty.
We suggest that circuit layout, number of connections, and gate types, may be of interest for future research into what exactly makes a circuit difficult to reverse engineer.

In addition to solution probability, we analyzed the distribution of solution times across tasks.
Interestingly, within Group~B, solution probability was not directly related to solution time -- tasks that were solved more frequently were not necessarily solved more quickly. 
This finding emphasizes the importance of considering both metrics when assessing participant performance.
Among the non-obfuscated tasks in groups~A to C, we observed that -- as the number of outputs increased from Group~A (one output) to Group~C (three outputs) -- both the mean solution time and its variance expanded by multiple minutes. 
This result suggests that as the problem space becomes more complex, identifying relevant components and devising an efficient traversal strategy become crucial factors that heavily influence solution time.
Lastly, we observed that reverse engineering the obfuscated circuits in Group~D yielded solution probabilities and times that are between those of groups B and C.
Introducing a single obfuscated gate does not appear to increase the problem space more than the addition of a third output.

\subsection{Discussion of RQ3: \reversim Integrates Cognitive Tests, Reveals Correlations with Task Performance}
\label{section:pilot:discussionrq3}

\paragraph{RQ3a: Integrating Standardized Cognitive Tests in \reversim}
The successful integration of the number-connection test into \reversim is supported by minimal participant dropout and solution times for matrices 1-3 that align with the norm values from the test manual for our well-educated sample.
The significant deviation observed in Matrix~4, which was not reported in the test manual, likely results from participant fatigue.
This observation suggests that potential fatigue effects should be considered when adapting similar tests to unsupervised online environments.
Given these considerations, we conclude that the number-connection test in \reversim is suitable for its intended purpose, \ie, measuring cognitive processing speed.
Furthermore, incorporating other standardized cognitive tests, such as the n-back task for working memory, appears feasible for evaluating additional cognitive factors relevant to \ac{HRE}.

\paragraph{RQ3b: Insights into the Human Aspects of \ac{HRE}}
Our findings indicate that integrating standardized cognitive tests into \reversim offers valuable insights into the human aspects of \ac{HRE}.
For instance, we observed a small correlation between processing speed and solution time, suggesting that individuals who process information faster tend to solve the \ac{HRE} tasks more quickly. 
Additionally, we found a moderate correlation between processing speed and the total number of tasks solved.
While it might be tempting to attribute this to faster thinkers having more time for the complex tasks, we observe a very similar correlation in those participants who did not reach the 75-minute time limit.
Hence, cognitive processing speed, as a component of intelligence, likely mediates other cognitive processes such as learning and reasoning.
These exploratory findings, which emerged from the initial implementation of the tests, align with existing research on the role of cognitive factors like working memory in \ac{HRE} performance~\cite{becker2020exploratory}. 
Although systematic studies are needed, these results highlight the potential of integrating cognitive tests to further investigate the cognitive dimensions of \ac{HRE}.
Identifying cognitive factors that correlate particularly strongly with \ac{HRE} performance could highlight promising targets for cognitive obfuscation, which could place greater demands on these factors and potentially lead to more comprehensive security-relevant conclusions in the future.

\section{Human Aspects of \texorpdfstring{\acs{HRE}}{HRE} in Novices and Intermediates: Similarities and Differences}
\label{section:intermediate}
\begin{figure}[t]
    \centering
    \includegraphics[width=.95\columnwidth]{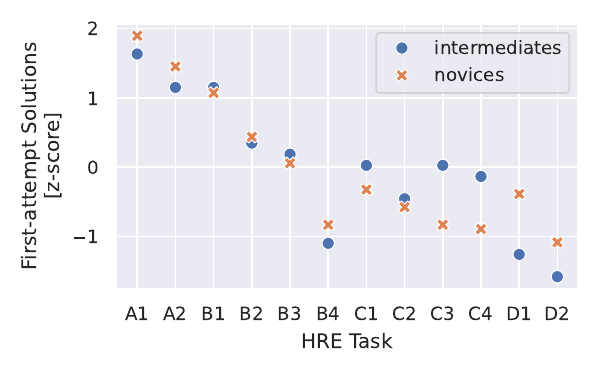}
    \caption{$z$-standardized number of participants solving a task on first attempt.\protect\footnotemark{} The plot highlights the relative performance differences \textit{between} tasks, for novices and intermediates, correcting for the fact that intermediates tend to perform better overall. For tasks A1 to C2, data from the intermediate sample closely tracks the novice sample. Tasks C3 and C4 appear relatively \textit{easier} for intermediates, while -- surprisingly -- tasks D1 and D2 appear \textit{harder} for them.}
    \Description[plot of the z-score for first-attempt solutions over every HRE task, divided into intermediate and novice participants]{Scatter plot showing the number of first-attempt solutions for all 12 HRE tasks, split by intermediates and novices. Due to the z-score, the points for intermediate and novice are close together for each task, and the number of first-attempt solutions can be approximated with a trend line ranging from 2 down to -1.5. However, there are 3 notable exceptions: B4 is an outlier with -1 as the z-score for intermediates and novices, indicating this task was easy to solve on the first try. As mentioned in the caption, intermediates struggled with C3 and C4, while novices were close to the trend line. This condition flips for D1 and D2, where intermediates stay close to the trend line and novices deviate.}
    \label{fig:results:zscores}
\end{figure}

To identify which insights \reversim may provide about human aspects of \ac{HRE} beyond novices, it is important to understand how behavior changes with participants more experienced in \ac{HRE}.
Thus, we repeated our user study with 47 more experienced participants, largely composed of computer science students and attendees of a workshop on \ac{HRE} from both academia and industry.
This intermediate group reported a mean prior knowledge of $M = 2.67~(SD = 0.83)$, corresponding to between \enquote{low} and \enquote{medium}.
In our first study with novices, in contrast, participants had reported a mean prior knowledge of \enquote{very low} $(M = 1.10)$. 
\autoref{appendix:intermediatesample} contains the full details of our sample and the results.
To compare the performance of novices and intermediates, we analyzed the proportion of participants who solved each task correctly on first attempt.
Our findings reveal both strong similarities and two distinct differences, which we present in \autoref{fig:results:zscores}.
Each provides insight into human aspects of in \ac{HRE}, such as how expertise develops and affects cognitive strategies in \ac{HRE}, and encourages further research with \reversim.

\footnotetext{A $z$-score of 1 indicates that the number of participants who solved the task on first attempt is one standard deviation above the sample mean.}

As expected, the intermediates performed better overall, reflecting their higher baseline knowledge and skills. 
To examine whether their performance advantage varied across different tasks,
we subsequently controlled for this overall performance difference by $z$-standardizing the data.
Specifically, for both samples we calculated $z$-scores by subtracting the sample mean from each data point and dividing by the standard deviation.
Doing so enables us to compare the performance of novices and intermediates in each task in terms of standard deviations.
Our findings show that for the eight simpler tasks (A1 to C2), the relative performance difference between the samples remained small, with variations not exceeding half a standard deviation.
This result suggests that while intermediates performed better in general, the expertise they have gained did not provide a distinct advantage in these specific tasks.
Instead, the growing complexity of these tasks appears to similarly decrease the performance of novices and intermediates.
For the most complex tasks (C2 to D2), however, we observe an interesting divergence in performance trends: While novices show a steady decline in performance for tasks in Group~C as complexity increases, intermediates maintain nearly constant performance.
Surprisingly, this effect is reversed in Group~D tasks involving obfuscated gates, where novices perform better than intermediates in relative terms.

This observed divergence may reflect underlying differences in how novices and intermediates approach these specific tasks, offering valuable insights into the process of  \ac{HRE} expertise acquisition.
Intermediates may have developed advantageous skills such as advanced pattern recognition or domain-specific heuristics, which help reduce cognitive load when encountering known structures.
However, these domain-specific approaches may be less effective, or even detrimental, when encountering unexpected or ill-formed circuits.
Novices, lacking these ingrained strategies, may approach such tasks with generally slower but more flexible or exhaustive reasoning methods.
Going forward, identifying the task-specific features that cause such performance differences within \reversim may help researchers to develop evidence-based prototypical mechanisms for cognitive obfuscation.

\section{Overall Discussion \& Conclusion}

\acf{HRE} plays a vital role for increasing trust in \acp{IC}, which is a primary objective of recent US and EU microchip initiatives~\cite{repryan20224346,eu2023regulation}.
As \ac{HRE} cannot be fully automated, human problem-solving competencies and cognitive abilities are critical to its effective execution~\cite{becker2020exploratory,wiesen2021anatomy}.
To enable controlled and quantitative studies of these human aspects in \ac{HRE}, which remain poorly understood~\cite{fyrbiak2017hardware}, we developed \reversim, an \ac{HRE} study environment that integrates standardized, unsupervised cognitive tests.
We evaluated \reversim with experts and non-experts across three research questions, summarized below, and make it available under an open-source license to provide researchers with an effective method for studying the human aspects in \ac{HRE}.

\noindent\fcolorbox{gray!80}{gray!10}{%
    \parbox{\columnwidth-30pt}{%
        \paragraph{Summary of RQ1}
        Experts confirmed that \reversim captures key aspects of real-world netlist reverse engineering, particularly in replicating problem-solving strategies and circuit analysis challenges.
        They also identified opportunities to extend \reversim's functionality, such as incorporating more complex logic gates and sequential elements, to better reflect the full complexity of netlist analysis.
    }
}\\[5pt]

\noindent\fcolorbox{gray!80}{gray!10}{%
    \parbox{\columnwidth-30pt}{%
        \paragraph{Summary of RQ2}
        \reversim demonstrates its suitability for studying aspects of \ac{HRE} with non-experts, as participants with minimal prior knowledge were able to engage meaningfully with the tasks. 
        The environment effectively differentiates participant performance across a wide range of task difficulties, indicating that it is well-suited for assessing \ac{HRE} strategies in a non-expert population.
    }
}\\[5pt]

\noindent\fcolorbox{gray!80}{gray!10}{%
    \parbox{\columnwidth-30pt}{%
        \paragraph{Summary of RQ3}
        Our findings demonstrate that \reversim can be effectively used to study cognitive factors related to \ac{HRE} performance by integrating standardized cognitive tests, such as the number-connection test. 
        These integrations offer insights, such as the correlation between cognitive factors and task performance, highlighting \reversim's potential for exploring the human aspects in \ac{HRE} in future research.
    }
}\\[5pt]

While our contributions to this complex area of research are notable, they come not without limitations:
First, we designed \reversim from scratch to enable controlled experimental studies that can accurately assess the impact of cognitive factors on \ac{HRE} problem-solving processes.
While \reversim cannot capture all aspects of real-world \ac{HRE}, our interview study suggests that it effectively models key processes and thought patterns, while also laying the foundation for implementing other important \ac{HRE} subprocesses, such as sequential circuit analysis and module recognition.
Although our current work does not yet provide far-reaching insights into \ac{HRE} countermeasures, such as the use of obfuscation primitives, the development and empirical evaluation of \reversim are essential steps in advancing this field.

Second, in all studies, we asked participants to self-rate their prior knowledge.
Although we have no evidence that participants \textit{systematically} over- or underestimated their prior knowledge, these self-assessments may be prone to bias and therefore may not accurately reflect participants' actual knowledge.
In our interview study, where a high level of \ac{HRE} expertise was essential, we therefore included additional metrics to assess the prior knowledge of the interviewees.
For future studies aimed at assessing the cognitive factors that influence \ac{HRE} success, further evidence-based determination of prior knowledge, beyond experience in computing or microchips, may be required. 

Third, we observed a significant performance loss in Matrix~4 of the number-connection test in the user study with novices.
While meta analyses have shown no differences between computerized and paper-and-pencil cognitive tests~\cite{Mead1993}, solving matrices on a screen may be more cognitively demanding.
Fatigue or loss of motivation, more likely in an unsupervised online setting, could also contribute to this decline.
Future studies should assess  participants' fatigue and motivation between matrices to better understand the performance drop.
Additionally, unlike the paper version, our digital implementation does provide feedback for misclicks, which may affect performance.
While individual performance should be interpreted cautiously, our implementation aimed to compare differing cognitive processing speeds rather than provide psychological diagnoses.
For this purpose, an exact match to the pen-and-paper version is not required, and we consider the test applicable.

\paragraph{Outlook}
Below we briefly outline research avenues enabled by \reversim as a controlled environment:
In future studies, \reversim can be used to empirically determine which characteristics of a circuit affect the cognitive load of reverse engineering it, including those that remain challenging even with increasing experience.
In this way, \reversim can support the development of design patterns for \enquote{cognitively obfuscated} \acp{IC}~\cite{wiesen2019towards}.
As a second line of defense to traditional obfuscation techniques, this type of hardware protection aims to create circuitry that deliberately slows down a human attacker's problem-solving process.
To this end, it seems worthwhile to examine whether concepts from software obfuscation~\cite{collberg1997taxonomy}, such as \textit{potency} (degree of human confusion) and \textit{resilience} (required time and effort to automate analyses), can be transferred to the realm of hardware obfuscation.
Conversely, understanding how essential \ac{HRE} skills are acquired and where problem solvers struggle may also inform the development of a training environment and educational guidelines for hardware security analysts.
Hence, \reversim may be a useful tool in meeting the growing need for hardware security professionals.
By enabling studies with larger and more diverse sample sizes, \reversim allows researchers to validate strategies and cognitive factors identified in prior studies~\cite{becker2020exploratory,wiesen2021anatomy} and uncover how additional cognitive factors and growing expertise influence \ac{HRE} performance, addressing previous research limitations.
We emphasize that it is important to complement quantitative results obtained with \reversim by validating them, if possible, with \ac{HRE} experts in a more complex real-world environment such as HAL~\cite{fyrbiak2019hal}.

\begin{acks}
We are very grateful to Zehra Karadag, Max Hoffmann, and Jannik Schmöle for their support in the development of \reversim.
A big thanks also goes to Malte Elson, and Franziska Herbert for discussions and recommendations on our analysis methods, and to Julian Speith for project support.
Last but not least, we would like to thank all expert and non-expert participants for their time and effort invested in our study.
%This work was supported by the PhD School \enquote{SecHuman -- Security for Humans in Cyberspace} by the federal state of NRW, Germany and by Deutsche Forschungsgemeinschaft (DFG, German Research Foundation) under Germany´s Excellence Strategy -- EXC 2092 CASA -- 390781972.
This work was supported by the \grantsponsor{sechuman}{PhD School \enquote{SecHuman -- Security for Humans in Cyberspace} by the federal state of NRW, Germany}{https://sechuman.ruhr-uni-bochum.de/}, the \grantsponsor{dfg}{German Research Foundation (DFG)}{https://www.dfg.de/en} under Germany's Excellence Strategy - \grantnum{dfg}{EXC 2092 CASA - 390781972}, and the \grantsponsor{rctrust}{Research Center Trustworthy Data Science and Security}{https://rc-trust.ai} (\url{https://rc-trust.ai}), one of the Research Alliance Centers within the UA Ruhr (\url{https://uaruhr.de}).
% + Max H., Zehra, Jannik, all participants and especially experts; Markus, Malte, Franziska
\end{acks}

\bibliographystyle{ACM-Reference-Format}
\bibliography{bibliography}

\appendix
\section{\reversim Software and Study Materials}
\label{appendix:materials}
Digital artifacts such as the \reversim source code, binaries, and setup documentation are accessible at \url{https://github.com/emsec/reversim} under an open-source license, including the study settings introduced in \autoref{section:interview:methods} and \autoref{section:pilot:methods}.
Additionally, we provide the complete questionnaires and interview guide used in both studies at \url{https://osf.io/vcuyg/}.

\section{Detailed Codebook}
\label{appendix:detailedcodebook}
The final codebook used to annotate the content of each interview is presented below. 
The codebook is divided into three parts that reflect the three question blocks defined in the interview guide. 
In the following sections we present the \textit{categories} and \textbf{open codes} for each part, and provide a short description for each code. 
Part 3 was not further divided into categories and instead contains open codes at the top level. 

\subsection{Part 1: General Impression}
\paragraph{Positive Feedback}
\begin{itemize}
\item \textbf{Fun to participate}: Participant stated that taking part in \reversim was fun. 
\item \textbf{Obfuscation particularly interesting}: Participant noted that the obfuscated tasks were interesting to solve. 
\item \textbf{Annotation tools helpful}: Participant mentioned that it was helpful to use the drawing tools to solve the tasks. 
\item \textbf{Didactically well-structured}: Participant said that the interactive tutorial was didactically well-structured. 
\item \textbf{Intuitive design}: Participant stated that they felt well guided through the environment.  
\end{itemize}
\paragraph{Constructive Criticism}
\begin{itemize}
\item \textbf{Provide additional support}: Participant would like to have more support in solving the tasks, especially for non-expert participants. 
\item \textbf{Redesign obfuscated tasks}: Participant suggested redesigning obfuscated tasks as the obfuscated element was not in the critical analysis path and could be ignored. 
\item \textbf{Comparable strategies for analyzing obfuscated and unobfuscated tasks}: In order to solve the obfuscated tasks, the participant did not change their strategy they applied for solving the unobfuscated tasks. 
\item \textbf{Varying difficulties of interactive tutorial and main tasks}: Participant stated that the difference in complexity between the interactive tutorial and the main tasks was substantial. 
\end{itemize}

\subsection{Part 2: Comparing \reversim with Reality}
\paragraph{Elements Recognized from Real-World Netlist Reverse Engineering}
\begin{itemize}
\item \textbf{Found similar obfuscation schemes in real netlists}: Participant identified obfuscation principles (\eg, covert gates) in the environment that they knew from real netlists. 
\item \textbf{Switches}: Participant recognized switches in the tasks as (global) inputs from real netlists.
\item \textbf{Wires}: Participant stated that the electrical connections in a netlist are comparable to the wires depicted in the tasks. 
\item \textbf{Outputs}: Participant said that the outputs of the circuits in the tasks (\ie, lamp, danger sign) are comparable to the concept of (global) outputs in real netlists. 
\item \textbf{Logic zero and one}: Participant considers the concept of binary values in tasks comparable to that in real (digital) netlists.
\item \textbf{Basic combinational gates}: Participants identified basic combinational gates (\ie, AND, OR, NOT) that they also find in real netlists.
\end{itemize}
\paragraph{Real-World Elements Missing in \reversim}
\begin{itemize}
\item \textbf{Specific basic elements missing}: Participant said that specific basic elements (\eg, NAND, NOR, XOR) were not included in the environment but do occur in real netlist analysis.
\item \textbf{Overall complexity of circuits too low}: Participant stated that overall complexity of the tasks in \reversim in too low in comparison to real \ac{HRE}. 
\item \textbf{No sequential logic}: Participant mentioned that in real-world netlist analysis, they also analyze sequential logic that was not part of the tasks. 
\item \textbf{Obfuscated gates never seen in reality}: Participant noted that they have never seen obfuscated gates in real netlists. 
\end{itemize}
\paragraph{Real-World Approaches Covered by \reversim}
\begin{itemize}
\item \textbf{Analysis of modules:} Participants recognized parallels between the analysis of \reversim's tasks and the analysis of (sub)modules in real netlists. 
\item \textbf{Annotation, pen and paper:} Participant stated that they used the drawing tools to make annotations and recognized parallels to the annotation process in real netlists. 
\item \textbf{Output-driven approach, back justification:} Participant stated that they solved the tasks by focusing on the outputs and said that this approach is also common in real \acsp{HRE}.
\item \textbf{Hypothesis-driven approach}: Participant formulated a hypothesis about an element or an input-output behavior and started to analyze the circuit based on this hypothesis. Furthermore, the participant said that this approach is comparable to real \ac{HRE}. 
\item \textbf{Simulation:} Participant said that they simulated specific outputs of the circuit (\eg, with the drawing tools) which is comparable to key processes of simulation in real netlists.  
\item \textbf{``I saw the solution'':} Participant said that they just saw the correct solution for a task what may also occur in real netlists with low complexities. 
\end{itemize}
\paragraph{Real-World Approaches Missing in \reversim}
\begin{itemize}
\item \textbf{No semi-automated tools:} Participant said that the usage of semi-automated tools is a common practice in \ac{HRE} but missing in the environment. 
\item \textbf{No (automated) dynamic analysis:} Participant mentioned that dynamic analysis (\ie, simulation, which, in this context, refers to the visualization of current flow in a netlist) is commonly used in netlist analysis, but was not mapped in \reversim.
\item \textbf{No modularization of netlists:} Participant noted that modularization of netlist elements is a common practice in real \ac{HRE} but was not included in the environment. 
\item \textbf{Specific solution approaches missing:} Participant stated that a few specific solution approaches they usually apply in \ac{HRE} (\eg, input-driven approach, trial and error, brute-force) were not part of the environment. \end{itemize}
\paragraph{Evaluation of Interaction Mechanics}
\begin{itemize}
\item \textbf{Rules force you to think}: Participant said that the interaction mechanics and rules forced them to think and to apply specific strategies. 
\item \textbf{Brute-force approach effectively prevented}: Participant stated that the interaction mechanics successfully prevented brute-force approaches, as brute-force is inefficient in real netlists. 
\item \textbf{Rules feel natural}: Participant mentioned that the interaction mechanics forced them to apply strategies and procedures that they would also apply when analyzing real netlist, and therefore they did not feel unnatural in the \acsp{HRE} context. 
\end{itemize}

\subsection{Part 3: Future Research \& Additional Features}
\begin{itemize}
\item \textbf{Add further combinational gates:} Participant suggested adding further combinational elements such as NAND or XOR in future studies with the environment. 
\item \textbf{Add sequential logic:} Participant suggested adding sequential logic such as flip-flops in future studies with the environment. 
\item \textbf{Add high-level components:} Participant suggested adding high-level components such as adders in future studies with the environment. 
\item \textbf{Add waveforms:} Participant suggested adding waveforms in future studies with the environment. 
\item \textbf{Additional objectives for tasks:} Participant suggested adding objectives to the tasks, \eg, to determine gate functionality based on a presented circuit. 
\end{itemize}

\newpage
\onecolumn
\section{Description of the Interview Study Sample}
\label{appendix:experts}
\begin{table*}[h!]
    \centering
    \caption{Demographic data and information on the technical and \ac{HRE} expertise of the 14 interviewees.}
    \label{tab:expert_sample}
    \begin{tabular}{ccllccccccccc}
        \toprule
        ID & \multirow{2}{*}{\parbox{0.06\textwidth}{\centering Age group}} & \multirow{2}{*}{\parbox{0.07\textwidth}{Highest degree}} & \multirow{2}{*}{\parbox{0.07\textwidth}{Employer sector}} & \multirow{2}{*}{\parbox{0.07\textwidth}{\centering Years in \ac{HRE}}} & \multicolumn{2}{c}{Weekly hours for tasks} & \multirow{2}{*}{\parbox{0.085\textwidth}{\centering Systems analyzed}} & \multicolumn{3}{c}{Experience self-rating} \\\cline{6-7} \cline{9-11} 
           &           &                &                  &                       & \ac{HRE} & other technical && general & theor. & practical \\
        \midrule
          1 & 18-29 & Master   & ECON & 2 & 10-20 & 20-30 & 0-3    & 4 & 4.1 & 4.36 \\
          2 & 18-29 & Bachelor & --   & 3 & --    & 40    & 4-6    & 4 & 3.8 & 3.45 \\
          3 & 18-29 & Master   & --   & 4 & 20-30 & 10-20 & 7-10   & 5 & 4.7 & 4.64 \\
          4 & 30-39 & PhD      & IC   & 7 & 5-10  & 40    & 51-100 & 4 & 4.7 & 4.64 \\
          5 & 18-29 & Master   & --   & 5 & 20-30 & 10-20 & 11-25  & 4 & 4.1 & 4.09 \\
          6 & 18-29 & Master   & IC   & 4 & 10-20 & 10-20 & 7-10   & 3 & 3.3 & 3.00  \\
          7 & 40-49 & PhD      & IC   & 7 & 10-20 & 30-40 & 51-100 & 5 & 4.7 & 4.45 \\
          8 & 30-39 & Master   & IC   & 4 & 20-30 & 20-30 & 0-3    & 4 & 3.6 & 2.55 \\
          9 & 30-39 & PhD      & PUB  & 7 & 20-30 & 20-30 & 7-10   & 5 & 4.1 & 3.55 \\
         10 & 30-39 & Master   & IC   & 5 & 20-30 & 30-40 & 4-6    & 5 & 4.1 & 4.18 \\
         11 & 18-29 & Bachelor & --   & 2 & --    & 5-10  & 4-6    & 3 & 3.8 & 3.91 \\
         12 & 30-39 & Master   & MANU & 4 & --    & 5-10  & 4-6    & 3 & 2.9 & 2.00 \\
         13 & 18-29 & Master   & ECON & 4 & 30-40 & 5-10  & 11-25  & 4 & 3.7 & 3.73 \\
         14 & 18-29 & Master   & CS   & 3 & 20-30 & 10-20 & 4-6    & 4 & 3.0 & 3.91 \\
        \bottomrule
    \end{tabular}
    \caption*{
       \raggedright
       \begin{footnotesize}
       \begin{tabular}{lp{0.35\textwidth}lp{0.45\textwidth}}
            CS & Computer security & MANU & Manufacturing/production of goods, other industry \\
            IC & Information and communication & PUB & Public administration, defense, education, health care and social services \\
            ECON & Scientific/technical services, other economic services 
       \end{tabular}
       \end{footnotesize}
    }
    \Description[demographics and levels of expertise of the interview study participants]{The table on demographics and both \ac{HRE} and further technical expertise of the interviewees consists of 11 columns. The first column contains the participant ID. The following four columns contain the age group of participants with 10 years granularity, the highest university degree, abbreviation for the business sector of the employer, and years spent working in \ac{HRE}. They are followed by two columns indicating the weekly number of hours working on \ac{HRE} tasks, and other technical tasks. A further column indicates a range for the number of hardware systems analyzed. Finally, there are three columns for self-rating scores, divided into general \ac{HRE} experience, theoretical knowledge, and practical skills. There are 14 data rows. The abbreviations for the employer sector are described in the table footnotes.}
\end{table*}

\section{Items of Theoretical-Knowledge Scale and Practical-Skill Scale from the Interview Study}
\label{app:section:interviewsubitems}
\begin{table*}[h!]
    \centering
    \caption{Self-rating items for the interview study participants on a scale ranging from 1~(very low) to 5~(very high) }
    \begin{subtable}[t]{0.45\textwidth}
    \caption{Subitems of the theoretical-knowledge scale.}
    \label{tab:appendix:interviewsubitems:theoretical}
    \begin{tabular}{p{5cm}rrr}
    \toprule
     Items                                                  &  $M$ & $SD$ & $N$ \\
    \midrule
     Boolean Algebra                                        & 4.29 & 0.72 &  14 \\
     Integrated Circuits (IC)                               & 3.86 & 0.77 &  14 \\
     \acsp{FPGA}                                                  & 3.86 & 1.02 &  14 \\
     Hardware Design                                        & 3.57 & 0.85 &  14 \\
     Processes and Methods of Netlist Extraction from \acsp{IC}   & 4.50 & 0.85 &  14 \\
     Processes and Methods of Netlist Extraction from \acsp{FPGA} & 3.50 & 1.45 &  14 \\
     Methods of Netlist Analysis                            & 4.07 & 0.91 &  14 \\
     Intellectual Property Protection                       & 3.79 & 0.97 &  14 \\
     Hardware Obfuscation                                   & 3.86 & 0.77 &  14 \\
     Hardware Trojans                                       & 3.71 & 1.20 &  14 \\
    \bottomrule
    \end{tabular}
    \Description[list of subitems of the theoretical-knowledge scale, and descriptive statistics]{The table on the subitems of the theoretical-knowledge scale contains four columns. The leftmost column contains the actual item text as used in the questionnaire of the interview study. The following three columns indicate the mean, standard deviation, and sample size in the interview study for each item.}
    \end{subtable}
    \hspace{3em}
    \begin{subtable}[t]{0.45\textwidth}
    \centering
    \caption{Subitems of the practical-skill scale.}
    \label{tab:appendix:interviewsubitems:practical}
    \begin{tabular}{p{5cm}rrr}
    \toprule
     Items                                      &  $M$ & $SD$ & $N$ \\
    \midrule
     Netlist Extraction from \acsp{IC}                & 3.86 & 1.23 &  14 \\
     Netlist Extraction from \acsp{FPGA}              & 3.14 & 1.46 &  14 \\
     Usage of \acs{HRE} Tools                         & 4.00 & 1.10 &  14 \\
     Reverse Engineering of Netlists            & 4.21 & 1.05 &  14 \\
     Detection of \acfp{FSM}  & 3.43 & 1.22 &  14 \\
     Detection of Specific combinational Blocks & 3.64 & 1.27 &  14 \\
     Analysis of Data Flow                      & 3.86 & 1.23 &  14 \\
     Netlist Simulation                         & 3.79 & 1.25 &  14 \\
     Object-Oriented Programming                & 4.07 & 0.91 &  14 \\
     Procedural Programming                     & 3.64 & 1.08 &  14 \\
     Hardware Description Languages             & 3.57 & 1.08 &  14 \\
    \bottomrule
    \end{tabular}
    \Description[list of subitems of the practical-skill scale, and descriptive statistics]{The table on the subitems of the practical-skill scale contains four columns. The leftmost column contains the actual item text as used in the questionnaire of the interview study. The following three columns indicate the mean, standard deviation, and sample size in the interview study for each item.}
    \end{subtable}
\end{table*}

\newpage
\onecolumn

\section{Task Parameters of the User Study}
\label{app:section:leveldetails}

\begin{table*}[h!]
    \centering
    \caption{Parameters for the four qualification tasks and the 12 tasks of the four complexity groups. The binary strings in the \textbf{Target Outputs} column describe the output types: A `1' stands for a lamp and a `0' stands for a danger sign, for example the string `101' stands for the three outputs lamp, danger sign, lamp (see~\autoref{fig:leveldesign:examplelevels:c}). Likewise, the binary strings in the \textbf{Solutions} column indicate the switch positions for the correct solution(s), from top to bottom. A `1' stands for a closed switch and a `0' stands for an open switch, for example the string `010' stands for the three switch positions open, closed, open (see~\autoref{fig:leveldesign:examplelevels:c}). In our study, participants could choose to skip a task after four unsuccessful attempts or after 3 minutes~(Group~A), 10 minutes~(Group~B), 12 minutes~(Group~C), or 15 minutes~(Group~D). These times were determined through our piloting.}
    \label{tab:appendix:leveldesign:parameters}
    \begin{tabular}{ll*{8}{c}}
    \toprule
    \multicolumn{2}{l}{\textbf{Task ID}} & \multicolumn{5}{c}{\textbf{Gates}} & \multirow{2}{*}{\parbox{0.09\textwidth}{\centering\textbf{Target Outputs}}} & \textbf{\# Outputs} & \textbf{Solutions} \\ \cline{3-7}
          && AND   & OR   & Inverters & Camouflaged & Total \\
    \midrule
     \multicolumn{2}{l}{\textbf{Qualification}}  &             &            &              &            &              \\
       &  1 &           2 &          0 &            0 &             0 &             2 &               1 &          1 & 111           \\
       &  2 &           1 &          1 &            1 &             0 &             3 &               1 &          1 & 010           \\
       &  3 &           0 &          2 &            2 &             0 &             4 &               0 &          1 & 110           \\
       &  4 &           1 &          2 &            0 &             0 &             3 &               0 &          1 & 000, 100      \\
     \multicolumn{2}{l}{\textbf{Low (Group A)}} \\
        & 1 &           1 &          1 &            1 &             0 &             3 &               1 &          1 & 001, 100, 101 \\
        & 2 &           1 &          1 &            1 &             0 &             3 &               1 &          1 & 000, 100, 010 \\
     \multicolumn{2}{l}{\textbf{Medium (Group B)}} \\
        & 1 &           3 &          3 &            3 &             0 &             9 &              00 &          2 & 000           \\
        & 2 &           1 &          4 &            4 &             0 &             9 &              00 &          2 & 101           \\
        & 3 &           2 &          3 &            2 &             0 &             7 &              11 &          2 & 011           \\
        & 4 &           3 &          3 &            2 &             0 &             8 &              10 &          2 & 110           \\
     \multicolumn{2}{l}{\textbf{High (Group C)}} \\
        & 1 &           3 &          5 &            4 &             0 &            12 &             001 &          3 & 011           \\
        & 2 &           2 &          5 &            5 &             0 &            12 &             100 &          3 & 100           \\
        & 3 &           7 &          3 &            8 &             0 &            18 &             100 &          3 & 110           \\
        & 4 &           7 &          3 &            6 &             0 &            16 &             011 &          3 & 100           \\
     \multicolumn{3}{l}{\textbf{Camouflaged Medium (Group D)}} \\
        & 1 &           2 &          3 &            2 &             1 &             8 &              01 &          2 & 111           \\
        & 2 &           2 &          3 &            2 &             1 &             8 &              01 &          2 & 010           \\
    \bottomrule
    \end{tabular}
\end{table*}

\newpage
\section{Example Tasks for Each Complexity Type}
\label{app:section:complexity}

\begin{figure*}[h!]
    \centering
    \begin{subfigure}[t]{0.42\textwidth}
        \centering
        \frame{\includegraphics[height=.25\textheight]{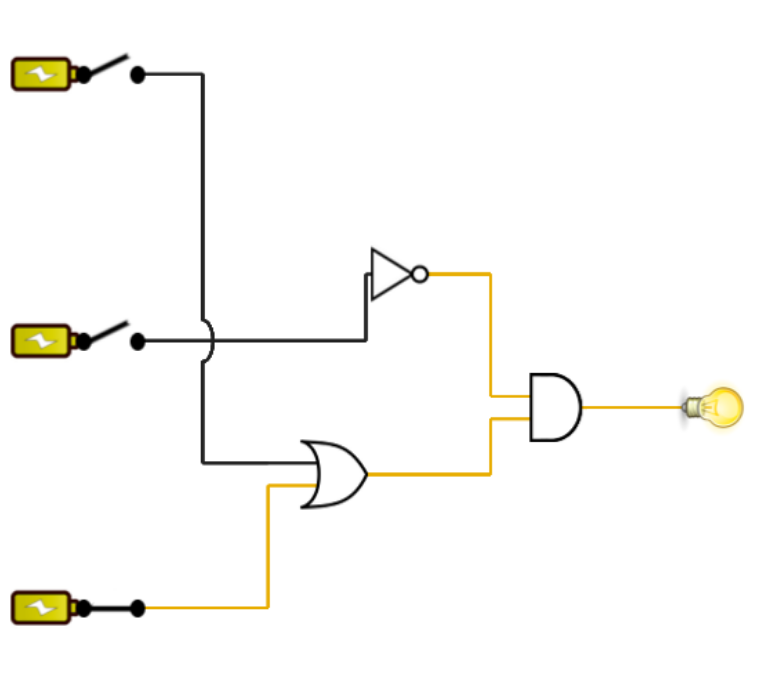}}
        \label{subfig:leveldesign:examplelevels:low}
        \caption{A low-complexity task features a single output and between 2 and 5 gates. All low-complexity tasks have three correct solutions.}
    \end{subfigure}
    \hfill
    \begin{subfigure}[t]{0.56\textwidth}
        \centering
        \frame{\includegraphics[height=.25\textheight]{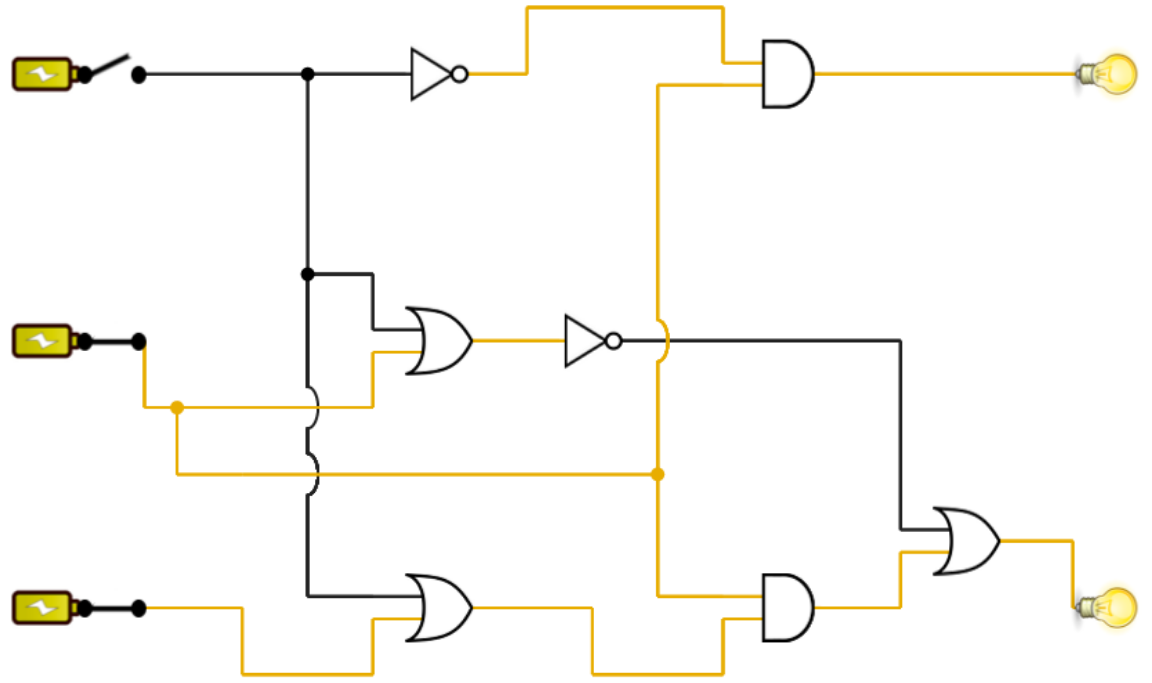}}
        \label{subfig:leveldesign:examplelevels:medium}
        \caption{A medium-complexity task features two outputs and between 7 and 11 gates. There is only one correct solution.}
    \end{subfigure}
    
    \vspace{.5cm}
    \begin{subfigure}[b]{\textwidth}
        \centering
        \frame{\includegraphics[height=.25\textheight]{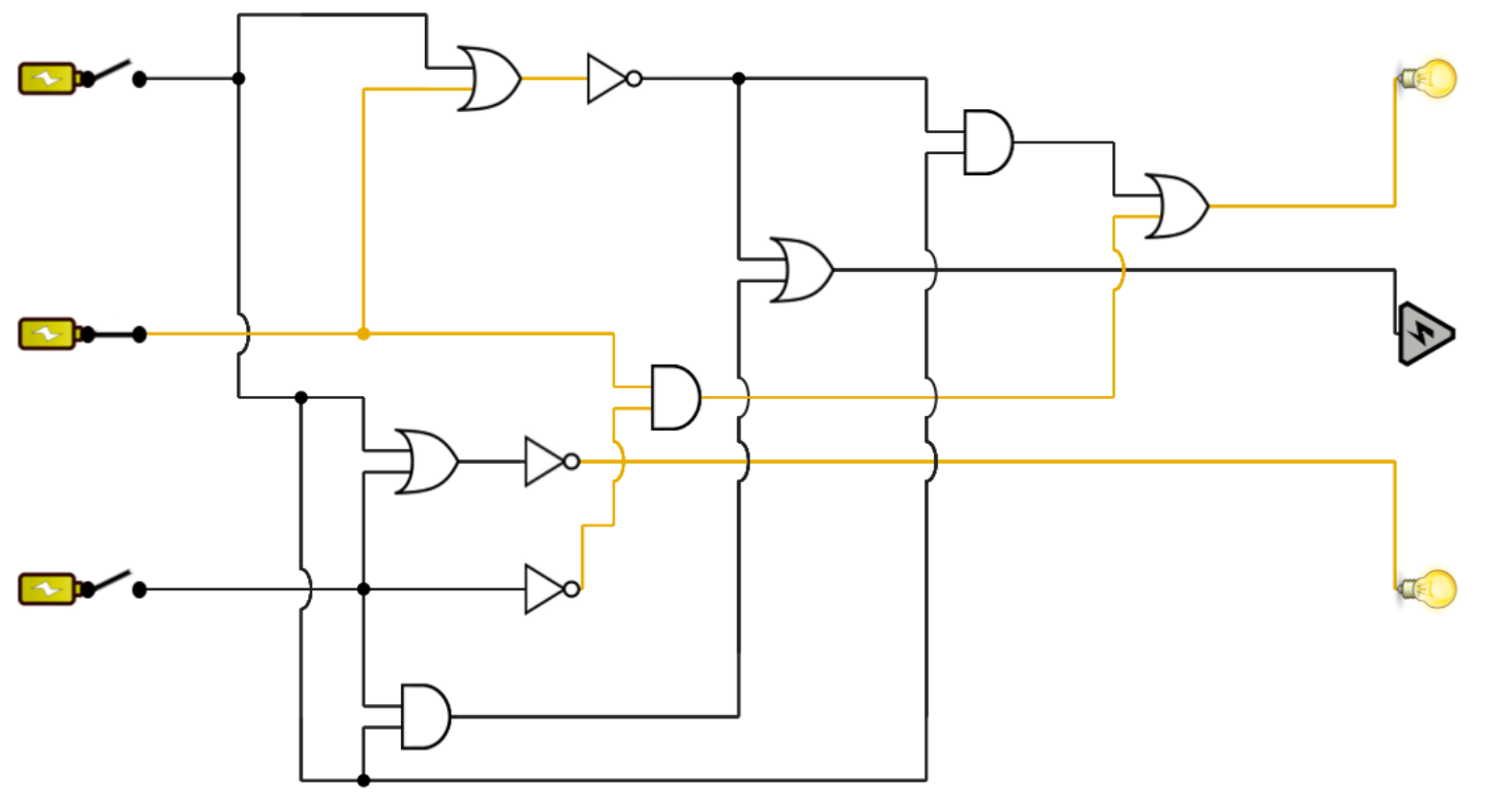}}
        \label{subfig:leveldesign:examplelevels:high}
        \caption{A high-complexity task features three outputs and between 10 and 18 gates. There is only one correct solution.}
    \label{fig:leveldesign:examplelevels:c}
    \end{subfigure}
    \caption{Tasks of increasing complexity differ in the size of the circuit and the number of outputs. All tasks feature exactly three inputs and can contain light bulbs and danger signs as outputs. The three examples each show a correct solution. Current-carrying wires are highlighted in yellow, as they would be after the participants submitted a correct solution. The colors have been optimized for printing; see \autoref{fig:simulation:basicelements:gameui} for an example of the screen representation.}
    \Description[three circuit diagrams for low, medium, and high-complexity tasks]{There are three subfigures. Each shows a circuit diagram as represented in the environment. All three diagrams feature three pairs of batteries and switches stacked vertically. The switches are in the positions required for the correct solution, such that at the outputs all light bulbs are lit and all danger signs are extinguished. Wires that are carrying current are highlighted in yellow. The low-complexity task consists of an AND, OR, and NOT gate and a single light bulb as an output. The bottom switch is closed, while the others are open. The medium-complexity task is larger, containing two AND gates, three OR gates, and two inverters. There are two outputs; both are light bulbs. The bottom two switches are closed. The high-complexity task is the largest. It contains three AND gates, four OR gates, and three inverters. From top to bottom, the outputs are a light bulb, a danger sign, and another light bulb. The middle switch is closed.}
    \label{fig:leveldesign:examplelevels}
\end{figure*}

\clearpage
\clearpage
\clearpage

\section{Items of Prior-Knowledge Scale from the User Study}
\label{appendix:playabilitysubitems}
\begin{table*}[h!]
    \centering
    \caption{Subitems of the domain-specific prior-knowledge scale calculated for the user study participants ranging from 0~(none) to 5~(very high).}
    \label{tab:appendix:playabilitysubitems}
    \begin{tabular}{lrrr}
    \toprule
     Items                                         &    M &   SD &   N \\
    \midrule
     Transfer: Propositional Logic                 & 0.91 & 1.28 &  99 \\
     Transfer: Boolean Algebra                     & 1.17 & 1.5  &  99 \\
     Transfer: Argumentation Logic                 & 0.79 & 1.13 &  99 \\
     Transfer: Set Algebra                         & 1.19 & 1.41 &  99 \\
     Transfer: Switching Algebra                   & 0.9  & 1.3  &  99 \\
     General: Digital Electronics                  & 1.52 & 1.54 &  99 \\
     General: Algorithms                           & 1.74 & 1.49 &  99 \\
     General: IT Systems                           & 1.96 & 1.57 &  99 \\
     General: Flowcharts                           & 2.33 & 1.39 &  99 \\
     Concrete: Logic Gates                         & 1.08 & 1.45 &  99 \\
     Concrete: Digital Circuits                    & 0.89 & 1.24 &  99 \\
     Expert: Object-oriented Programming Languages & 1.16 & 1.55 &  99 \\
     Expert: Procedural Programming Languages      & 1.14 & 1.44 &  99 \\
     Expert: Hardware Description Languages        & 0.7  & 1.11 &  99 \\
     Expert: Software Reverse Engineering          & 0.7  & 1.22 &  99 \\
     Expert: Hardware Reverse Engineering          & 0.38 & 0.91 &  99 \\
    \bottomrule
    \end{tabular}
    \Description[list of subitems of the domain-specific prior-knowledge scale, and descriptive statistics]{The table on the subitems of the domain-specific prior-knowledge scale contains four columns. The leftmost column contains the actual item text as used in the questionnaire of the user study. The following three columns indicate the mean, standard deviation, and sample size in the user study for each item.}
\end{table*}

\newpage
\twocolumn
\section{User Study with Intermediates}
\label{appendix:intermediatesample}

To identify similarities and differences in the performance of novices and intermediates when solving \ac{HRE} tasks in \reversim, we repeated our user study with a sample of participants with intermediate knowledge of \ac{HRE}.
For this study, we followed the same procedure as for the previously introduced user study.
Hence, we forego an elaborate description of the procedure here and instead refer the reader to \autoref{section:pilot} for details.

\subsection{Participants}
\label{appendix:intermediatesample:participants}
We recruited a total of 62 participants from attendees of a specialized workshop on \ac{HRE} from both industry and academia, as well as cybersecurity students with an interest in \ac{HRE}, following the same eligibility criteria as for the main study.
We had to exclude one participant due to technical issues, and 14 dropped out, resulting in a sample of $n = 47$ participants.
Participants in this study spent a median of 53 minutes completing the study.
They did not receive compensation.

Of our final sample, 36 participants identified themselves as male, nine as female, and two preferred not to disclose.
Participants were between 18 and 52 years old, with a mean age of $M = 27 (SD = 7.3)$.
27 participants had a university degree, 15 had a high school degree or equivalent, and one participant preferred not to disclose.

\subsection{Results}
\label{appendix:intermediatesample:results}

\subsubsection{Participants' Prior Knowledge \& Experience}
Our 61 participants self-rated their prior knowledge with a mean of $M = 2.67 (SD = 0.83)$, corresponding to between \enquote{low} and \enquote{medium}.
54 participants reported practical experience in computing and 34 reported practical experience with microchips.

\subsubsection{Participant Engagement \& Performance}
Of our 61 participants, 47 -- or 77\% -- completed the study and 14 dropped out. 
\autoref{appendix:intermediatesample:participants:flowchart} shows that the majority of dropouts, nine, occurred during the experiment phase, while four participants dropped out in the number-connection test and one during the qualification phase.
From the 14 dropouts, 13 participants had a background in computing, and ten had experience with microchips.

\begin{figure}[ht]
    \centering
    \includegraphics[width=.95\columnwidth]{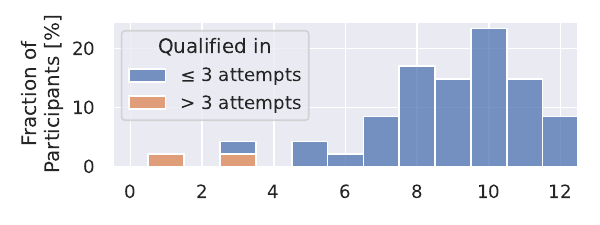}
    \caption{Number of \ac{HRE} tasks solved on the first attempt by fraction of participants, broken down into those requiring a maximum of three attempts to qualify and those requiring more than three attempts.}
    \Description[number of tasks solved first try by percent of players, divided into qualified in less than 4 attempts or 4 and more attempts]{histogram where each bar corresponds to the number of tasks that were solved in the first try, ranging from 0 to 12. The height of the bar represents the fraction of participants that solved this number of tasks. Each bar is divided into participants who qualified in under 4 attempts and participants who required 4 or more. The group that qualified in under 4 attempts resembles a normal distribution centered around 9 tasks, spanning the bars 5 to 12. Over 20 percent of this group solved 10 tasks on the first try. The group that qualified in 4 or more attempts makes up about 5 percent and solved between 1 and 3 tasks.}
    \label{appendix:intermediatesample:results:attempts_distribution}
\end{figure}

In the following, we consider the 47 participants completing the study, four of whom reached the time limit and missed a median of 2.5 tasks.
Participants solved a mean of $M = 11.4$ tasks with any number of attempts, and a mean of $M = 8.8$ tasks on first attempt.
Overall, find that 89\% of participants only took a single attempt to qualify.
\autoref{appendix:intermediatesample:results:attempts_distribution} shows that the two participants who required more than three qualification attempts solved one and three tasks on first attempt, respectively, while the 45 participants qualifying with three or fewer attempts solved a mean of nine tasks -- an even more pronounced difference than what was observed in the main study.

\subsubsection{Participants' Feedback}
As for the main study, none of the 47 non-dropout participants reported accessibility issues when using \reversim.
Participants agree or fully agree ($M = 4.51, SD = 0.59$) that they enjoyed engaging with \reversim, and agree that they understood the interaction mechanics ($M = 4.70, SD = 0.57$).
They were undecided whether the scoring was motivating ($M = 3.85, SD = 0.88$).
We observed bimodality in participants' opinion on the opportunity to repeat the tutorial with a major mode of 3 (\enquote{neither nor}) and a minor mode of 5 (\enquote{fully agree}), independently of whether participants did repeat the tutorial.
Participants fully agreed, however, that the tutorial was easy to understand ($M = 4.87, SD = 0.34$).
A total of 44 participants indicated having used the drawing tools, fully agreeing that they are useful ($M = 4.84, SD = 0.37$).

\subsubsection{Per-Task Performance}
\begin{figure}[t]
    \centering
    \includegraphics[width=.95\columnwidth]{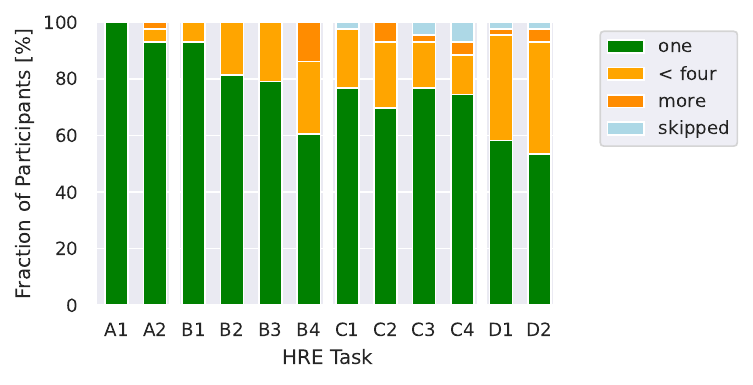}
    \caption{Fractions of participants who solved each task by the number of solution attempts.}
    \Description[Number of attempts for each task]{Stacked bar graph showing the number of attempts that were needed for each level. Each bar is divided into one solution attempt, less than 4 solution attempts, 4 or more than 4 solution attempts, and skipped. The tasks are A1-A2, B1-B4, C1-C4 and D1-D2. The number of first-try solutions steadily decreases from 100 percent for A1 down to 52 percent for D2. One strong outlier is B4, with only a 60 percent first-try solving rate and the highest amount of four or more attempts with around 15 percent. Another outlier is C3 and C4, with a first-try solving rate slightly above 75 percent, which is above the trend line. They also have the highest skip rate, at 5 to 10 percent.}
    \label{appendix:intermediatesample:results:attempts}
\end{figure}

To ensure comparability across tasks, we restrict our analysis to the 43 participants who did not reach the 75-minute time limit.
\autoref{appendix:intermediatesample:results:attempts} shows the fractions of participants who solved each task, broken down by the number of attempts.
If we ignore the number of attempts, then both Group~A and Group~B tasks were solved by 100\% of the participants, Group~C tasks by 93\% to 100\%, and Group~D tasks by 97\%.
If we only consider successful solutions in the first attempt, then the solution probabilities for Group~A tasks are 93\% to 100\%, for Group~B tasks 60\% to 93\%, for Group~C tasks 70\% to 77\% and for Group~D tasks 53\% to 58\%.

\begin{figure}[ht]
    \centering
    \includegraphics[width=.95\columnwidth]{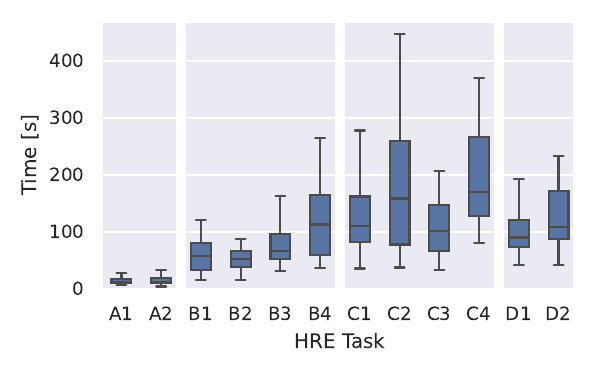}
    \caption{Time required by the participants to solve each \ac{HRE} task. Only the times of participants who solved each individual task correctly were included, \ie, sample size varies between 43 participants for Task A1 and 40 participants for Task C3.}
    \Description[mean time required by intermediates to solve each task]{The box plot shows the time participants spent on each task that they solved correctly. The x-axis shows the task names and the y-axis shows the number of seconds required to solve the task. For tasks A1-A2, participants took under 25 seconds to solve with a mean well below 50 seconds. For tasks B1-B4, participants took up to 270 seconds, with the mean time lying between 50 and 110 seconds. Tasks C1, C2, and C4 took up to 380 seconds, with mean times for tasks C1-C4 between 150 and 220 seconds. The upper whisker of C4 is exceptionally high at 440 seconds. Finally, for tasks D1-D2 up to 230 seconds were required to complete the tasks with a mean between 90 and 120 seconds. Overall, from group A to C, the tasks take longer to solve, while tasks in group D require a solution time between those of groups B and C.}
    \label{appendix:intermediatesample:results:timing_subset}
\end{figure}

\autoref{appendix:intermediatesample:results:timing_subset} shows the distribution of times to correct solutions for each \ac{HRE} task.
Median times were $12$ to $13$ seconds for Group A tasks, $52$ to $113$ seconds for Group B tasks, $101$ to $170$ seconds for Group C tasks, and $90$ to $109$ seconds for Group D tasks.

\begin{figure*}[t]
    \footnotesize
    \centering
    \begin{tikzpicture}[node distance=0.2cm and 0.7cm]
        \node[draw, align=center, thick] (presurvey) {Pre Survey\\($n=61$)};
        \node[draw, align=center, right = 4em of presurvey] (zvt) {Number\\Connection Test};
        \node[draw, align=center, right = of zvt] (tutorial) {Interactive\\Tutorial};
        \node[draw, align=center, right = of tutorial] (quali) {Qualification\\Phase};
        \node[draw, align=center, right = of quali] (competition) {Experiment\\with 12 Tasks};
        \node[draw, align=center, right = of competition] (end) {End};
        \node[draw, align=center, right = 4em of end] (postsurvey) {Post Survey\\($n=47$)};

        \draw[->] (quali) edge [dashed, bend right=50] node [midway, above, align=center] (requalifylabel) {repeat tutorial voluntarily\\or if not qualified ($8$)} (tutorial);
        \draw[->] (competition) edge [dashed, bend right=40] node [midway, below, align=center] (to2label) {timeout ($4$)} (end);

        \node[draw=gray, thick, inner xsep=1.5em, inner ysep=0.75em, fit=(zvt) (tutorial) (quali) (competition) (end) (requalifylabel) (to2label)] (ingame) {};
        \node[fill=white, align=center] at (ingame.north) {\reversim Environment};
        
        \node[align=center, below = 4.5em of quali] (dropout) {dropout (14)};
        
        \draw[->](presurvey) edge node[pos=.25, above] {61} (zvt)
            (zvt) edge node[midway, above] {57} (tutorial)
            (tutorial) edge node[midway, above] {57} (quali)
            (quali) edge node[midway, above] {56} (competition)
            (competition) edge node[midway, above] {43} (end)
            (end) edge node[pos=.70, above] {47} (postsurvey);

        \draw[->,densely dotted] (zvt) |- node[pos=.1, left] {4} (dropout);
        \draw[->,densely dotted] (quali) edge node[pos=.25, left] {1} (dropout);
        \draw[->,densely dotted] (competition) |- node[pos=.1, left] {9} (dropout);
    \end{tikzpicture}
    \caption{Overview of the flow of our user study with intermediates. Excluding one invalid dataset, 61 participants started the study. Eight revisited the tutorial at least once. 14 participants dropped out during the different phases of the study, particularly during the experiment phase, resulting in a total of 47 valid and complete datasets. Four participants did not finish within the 75-minute time limit before proceeding to the post survey.}
    \Description[flow of the intermediate user study]{Flowchart visualizing the progression of the intermediates through the study. The structure of the flowchart is the same as Fig. 6 but with different numbers. All participants start in "Pre Survey" and end up in either "Post Survey" or "dropout". All states except the start and exit states are inside the ReverSim environment. The primary flow direction is as follows: 61 participants move from "Pre Survey" to "Number Connection Test", 57 flow to "Interactive Tutorial", 57 flow to "Qualification Phase", 56 to "Experiment with 12 Tasks", 43 to "End" and 47 reach the "Post Survey". Additionally, 4 participants were redirected to "End" from "Experiment with 12 Tasks" via the timeout route. 8 participants flowed from "Qualification Phase" back to "Interactive Tutorial" because they did not qualify or wanted to repeat the tutorial. Out of the 14 dropouts, 4 came from "Number Connection Test", 1 from "Qualification Phase" and 9 from "Experiment with 12 Tasks".}
    \label{appendix:intermediatesample:participants:flowchart}
\end{figure*}

\subsubsection{Number-Connection Test Results}
57 participants completed all four matrices of the number-connection test.
Those participants solved Matrix~1 in a mean time of $M=57$ seconds (range: $35$ to $100$), the second in $M=57$ seconds (range: $39$ to $95$), the third in $M=57$ seconds (range: $33$ to $97$), and Matrix~4 in $M=66$ seconds (range: $36$ to $262$).
All distributions show a small positive skew.

A Welch's ANOVA~\cite{Welch1951} ($F = 1.41, df = 3, p = .24$) did not reveal a significant difference in means between all matrices.
Using matrices 1-4, the mean processing speed subcomponent of our participants' \acp{IQ} is about $116$, which is about one standard deviation above average (\ie, $100$).

\subsubsection{Correlation between Cognitive Processing Speed and \texorpdfstring{\acs{HRE}}{HRE} Task Performance}
\begin{figure}[ht]
    \centering
    \includegraphics[width=.95\columnwidth]{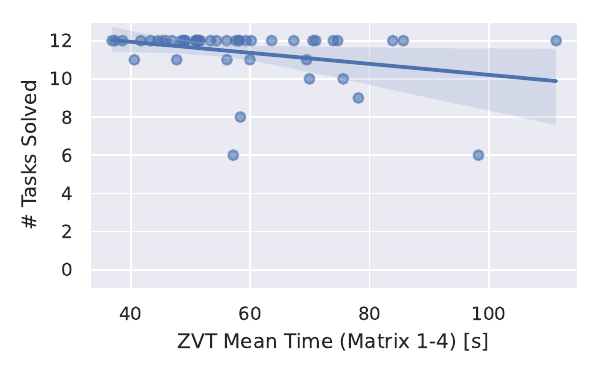}
    \caption{Mean time spent on the four matrices of the number-connection test (x-axis) and the number of tasks solved with any number of attempts (y-axis). We show the regression line to provide an intuition of the overall trend.}
    \Description[number of tasks solved over the mean time the intermediate participant took during ZVT]{The x-axis of the scatter plot shows the time in seconds that participants needed to complete the 4 tasks of the number-connection test. It ranges from 35 to 110 seconds. The y-axis shows the number of tasks solved, ranging from 0 to 12. All participants solved at least 6 tasks, while the vast majority solved 12 tasks in under 60 seconds. The few participants who solved less than 11 tasks all took at least 55 seconds to solve the ZVT. But a few outliers also took over 80 seconds to solve the ZVT and solved 12 tasks. The regression line shows a slight downward trend, meaning fewer tasks were solved by participants who took more time to complete the ZVT test. It ranges in a straight line from 12 tasks solved for 40 seconds to 10 tasks solved for 110 seconds.}
    \label{appendix:intermediatesample:results:zvt_corr}
\end{figure}

\begin{figure}[ht]
    \centering
    \includegraphics[width=.95\columnwidth]{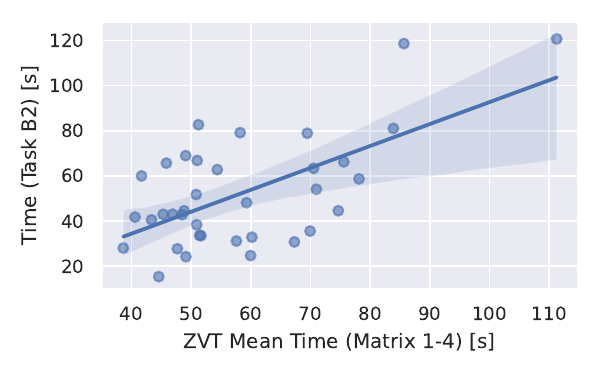}
    \caption{Mean time spent on the four matrices of the number-connection test (x-axis) and \textit{Time in Task}~B2, limited to those participants solving the task on first attempt (y-axis). We show the regression line to provide an intuition of the overall trend.}
    \Description[time in seconds for solving task B2 over the mean time the intermediate participant took during ZVT]{Scatter plot showing the time in seconds that intermediate participants needed to complete the number-connection test on the x-axis. It ranges from 35 to 110 seconds. The y-axis shows the solution time for all players who have finished task B2 on the first attempt. It ranges from 0 to 250 seconds. There is a high density of points in the bottom-left corner. The density of the point cloud decreases with increasing ZVT Mean Time. The regression line shows an upward trend, meaning participants who were quick in the ZVT were also faster in solving level B2 on the first try. It ranges in a straight line across the whole ZVT Mean Time range, starting at a y-intercept of 35 seconds and ranging to 105 seconds.}
    \label{appendix:intermediatesample:results:zvt_corrtime}
\end{figure}

Processing speed and the total number of \ac{HRE} tasks solved by each participant, visualized in \autoref{appendix:intermediatesample:results:zvt_corr}, are moderately correlated as indicated using Spearman's rank correlation ($\rho = -0.28$) and Kendall's $\tau$ ($\tau = -0.23$).
A Pearson correlation between processing speed and \textit{Time in Task}~B2 for participants solving the task on first attempt revealed a moderate positive correlation ($r = .62 $), visualized in \autoref{appendix:intermediatesample:results:zvt_corrtime}.
A similar Pearson correlation for processing speed and \textit{Time in Task}~B2 regardless of the number of attempts also showed a moderate positive correlation ($r = .42 $).
\cleardoublepage

\end{document}